\documentclass[11pt,a4paper]{article}

\usepackage{bbm}
\usepackage{bm}
\usepackage{jheppub}
\usepackage{mathrsfs}
\usepackage{mathtools}
\usepackage[dvipsnames]{xcolor}

\allowdisplaybreaks

\let\a=\alpha
\let\b=\beta
\let\g=\gamma
\let\d=\delta
\let\k=\kappa
\let\l=\lambda
\let\m=\mu
\let\n=\nu
\let\p=\phi
\let\r=\rho
\let\s=\sigma
\let\t=\tau
\let\c=\chi
\let\O=\Omega
\let\de=\partial

\renewcommand{\P}{\mathcal{P}}
\newcommand{\T}{\mathcal{T}}
\newcommand{\xx}{\hat{x}}
\newcommand{\EL}{\mathbbm{E}}
\newcommand{\E}{\mathcal{E}}
\newcommand{\bxi}{\bar{\xi}}
\newcommand{\bpsi}{\bar{\psi}}

\newcommand{\R}{\mathbbm{R}}
\newcommand{\un}{\mathbbm{1}}
\newcommand{\X}{\mathcal{X}}
\newcommand{\gr}[1]{\text{#1}}
\newcommand{\lie}[1]{\mathfrak{#1}}
\newcommand{\vek}[1]{\bm{#1}}
\newcommand{\dd}{\text{d}}
\newcommand{\D}{\text{D}}
\newcommand{\DD}{\mathbbm{D}}
\newcommand{\Da}{\text{D}}
\newcommand{\imag}{\text{i}}
\newcommand{\abs}[1]{\lvert#1\rvert}
\newcommand{\K}{\mathcal{K}}
\renewcommand{\L}{\mathscr{L}}
\newcommand{\mc}{\omega}
\DeclareMathOperator{\tr}{tr}
\DeclareMathOperator{\arctanh}{arctanh}

\newcommand{\boxedcolor}[2]{\color{#1}\boxed{\color{defaultcolor}#2}\color{defaultcolor}}

\colorlet{defaultcolor}{.}

\title{Geometric duality between effective field theories.\\
Part II. Exact classical solutions}

\author[a]{Tom\'a\v{s} Brauner,}
\author[b]{Diederik Roest,}
\author[b]{Tianzhi Wang,}
\author[a]{and Zhihao Zhang}

\affiliation[a]{Department of Mathematics and Physics, University of Stavanger,\\
4036 Stavanger, Norway}
\affiliation[b]{Van Swinderen Institute for Particle Physics and Gravity, University of Groningen,\\
9747 Groningen, The Netherlands}

\emailAdd{tomas.brauner@uis.no}
\emailAdd{d.roest@rug.nl}
\emailAdd{tian.wang@rug.nl}
\emailAdd{zhang.zhihao@uis.no}

\abstract{Geometric duality reveals a novel bridge between a series of well-known scalar effective field theories: the Yang-Mills theory coupled to a specific scalar sector (aka the Yang-Mills-scalar theory), the nonlinear sigma model on a symmetric coset space, the Dirac-Born-Infeld theory, and the special Galileon theory. In our previous work, we used the geometric duality to map scattering amplitudes in the various theories to one another. In this sequel, we show that the duality makes it also possible to transport exact classical solutions. Thus, exact solutions of the nonlinear sigma model are mapped to a subset of solutions of the Yang-Mills-scalar theory. Likewise, solutions of the special Galileon theory map to solutions of the Dirac-Born-Infeld theory, which in turn map to solutions of the nonlinear sigma model coupled to dynamical gravity.

For all these duality maps, we provide explicit examples of exact solutions of the source theory and their corresponding image in the target theory. While the duality maps are manifest at the level of certain geometric building blocks of the theories, we show that with a suitable choice of parametrisation, it is also possible to map directly the elementary scalar fields of the various theories to one another. This offers a practically convenient tool for constructing new exact solutions of the target theories.

Finally, the duality sequence including the nonlinear sigma model and the Yang-Mills-scalar theory can be extended by uplifting the solutions of the latter to a higher-dimensional spacetime. One thus arrives at a class of exact solutions of pure Yang-Mills theory in higher dimensions. In a similar vein, it is possible to obtain exact solutions of higher-dimensional pure gravity by uplifting solutions of the nonlinear sigma model, coupled to dynamical gravity, obtained via a duality map from solutions of the special Galileon theory.}

\keywords{Effective Field Theories, Classical Solutions, Minimal Surface}

\begin{document}

\maketitle


\section{Introduction}
\label{sec:intro}

Understanding the structure of quantum field theory, and the relations between different field theory models, remains one of the most important challenges in high energy physics. On this front, the recent decades have seen tremendous progress thanks to modern amplitude methods, drawing motivation from phenomenology and inspiration from string theory. The development of on-shell techniques has on the one hand pushed the frontier of scattering amplitude computations, and on the other hand led us to ask deep questions about the very nature of quantum field theory.

One of the most remarkable results of the scattering amplitude program has been the discovery that amplitudes in vastly disparate theories can be intimately related. This phenomenon has been dubbed \emph{double copy}; see refs.~\cite{Bern2019a,Bern2022,Adamo2022}. There are now several mutually complementary approaches that address relations between amplitudes across different theories, based on a cubic-vertex representation of Feynman diagrams~\cite{Bern:2010ue}, scattering equations~\cite{Cachazo2014a}, and combinatorics~\cite{Arkani-Hamed:2023mvg}. In parallel to the study of scattering amplitudes, substantial effort has been invested in understanding relations between classical solutions of different field theories, a property that goes under the name \emph{classical} double copy~\cite{Monteiro:2014cda}. Here, comparatively less is known than for amplitudes. Established results concerning exact solutions are mostly restricted to special classes of field configurations that make the matching of solutions of different theories possible.

In a prequel to this paper~\cite{Brauner:2025rzv}, we put forward a new type of duality that allows direct mapping of scattering amplitudes across a set of well-known effective theories of massless scalars including the Yang-Mills-scalar (YMS) theory, the nonlinear sigma model (NLSM), the Dirac-Born-Infeld (DBI) theory, and the special Galileon (sGal) theory, the latter three describing Nambu-Goldstone (NG) bosons. Our approach was inspired by earlier works utilizing classical equations of motion (EoMs) for the computation of scattering amplitudes~\cite{Monteiro2011,Cheung2021}. It however differs from these works substantially in that the underlying reason for the very existence of the duality map can be connected directly to the geometry of the individual theories in question. Using suitably defined geometric building blocks, the EoMs of different theories can be rewritten in a structurally equivalent form save for the interpretation of the field variables entering the equations. We thus end up with maps from a single source theory to a single target theory. Such relations, differing from the more traditional double copy where input from two theories is combined to produce a third theory, have been called ``doppelg\"anger'' in ref.~\cite{Moynihan:2025vcs}.

\begin{figure}
\centering
\includegraphics[width=0.75\textwidth]{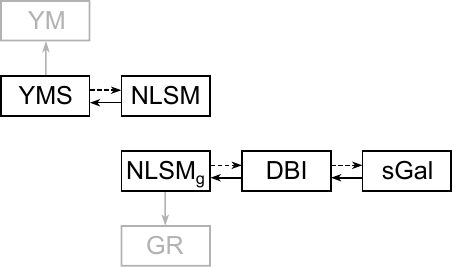}
\caption{Overview of the two duality sequences addressed in this paper; NLSM\textsubscript{g} refers to the nonlinear sigma model coupled to dynamical gravity. Importantly, the duality maps are directional. Scattering amplitudes of asymptotic single-particle states can be mapped to each other in the direction indicated by the dashed arrows; this case was analysed in our previous paper~\cite{Brauner:2025rzv}. The present paper deals with duality between exact solutions of the classical equations of motion, which are mapped to each other in the direction of the solid arrows. The horizontal alignment of the different theories is organized according to the soft degree that controls the scaling of scattering amplitudes in the soft limit. The uplift of exact solutions to pure Yang-Mills theory and pure gravity, discussed in section~\ref{sec:YM}, is indicated in grey.}
\label{fig:duality}
\end{figure}

Importantly, the geometric duality maps are directional. The (tree-level) scattering amplitudes are mapped in the direction indicated in figure~\ref{fig:duality} by the dashed arrows. In the present paper, we show that the same framework can be used to also map exact classical solutions, only the direction of the maps is the opposite (solid arrows in figure~\ref{fig:duality}). Our main result is that, in contrast to earlier works relying on solutions with a specific algebraic structure, the geometric duality can be used to map ``any'' exact solution of the source theory to a solution of the target theory.\footnote{Disclaimer concerning the quotation marks. As we will see later, the duality map involves reinterpreting spacetime indices in the source theory as flavour indices in the target theory. Requiring that the energy density of the target theory is bounded from below, we will thus only be able to map static (time-independent) solutions to static solutions. Alternatively, switching to Euclidean spacetime makes it possible to map truly all solutions.} It should be stressed that the relation between classical solutions of two theories obtained from the geometric duality is one of inclusion rather than equivalence. Namely, in all specific cases of duality, we will demonstrate that solutions of the source theory can be reinterpreted as a subset of solutions of the target theory, subject to certain additional relations that are not native to the latter theory.

The paper is organised around two separate duality sequences. In section~\ref{sec:EoMoverview1}, we outline a covariant formulation of the NLSM and the YMS theory, demonstrating the duality between the two. This is complemented in section~\ref{sec:Sol_Map1} by explicit examples of exact solutions of the NLSM and their images within the YMS theory under the duality map. Similarly, in section~\ref{sec:EoMoverview2}, we briefly review the construction of the sGal theory, the DBI theory, and the NLSM coupled to dynamical gravity (NLSM\textsubscript{g}). In section~\ref{sec:Sol_Map2}, we work out explicit sample solutions of these theories. As a bonus, we show in section~\ref{sec:YM} that the solutions of the YMS theory obtained from geometric duality can be uplifted to solutions of a higher-dimensional pure Yang-Mills (YM) theory, and that solutions of the NLSM\textsubscript{g} obtained from the sGal theory via geometric duality can be uplifted to solutions of higher-dimensional general relativity (GR). These uplifts are indicated in figure~\ref{fig:duality} in grey.

\paragraph{Note} This paper is a sequel to ref.~\cite{Brauner:2025rzv}. Unless stated otherwise, we adopt the same conventions as in that work. Further details on the constructions of the individual theories, including details specific to the duality between scattering amplitudes, can also be found there.


\section{Equations of motion for the NLSM-YMS duality sequence}
\label{sec:EoMoverview1}

In this section, we outline the geometric duality that embeds the classical solution space of the NLSM into that of the YMS theory. The duality map relies crucially on a covariant formulation of the NLSM, in which the geometric data characterizing the coset space of the NLSM are reinterpreted as the elementary propagating fields of the YMS theory. In order to make the building blocks entering the duality more transparent, we will highlight using a \textcolor{Green}{green box} equations that constitute the starting point (source) of the duality map. Likewise, we will highlight using a \textcolor{Red}{red box} equations that constitute the target of the duality map. The same notation is used below when we discuss the sGal-DBI-NLSM\textsubscript{g} duality sequence. The general rule of thumb is that in order to be able to claim that we have generated a solution of the target theory, we must successfully reproduce all its primary equations of motion (EoMs) along with any identities that the fields of the target theory might satisfy.


\subsection{Nonlinear sigma model}
\label{sec:NLSM}

Our starting point is the NLSM, which describes the dynamics of Nambu-Goldstone (NG) bosons associated with spontaneous breaking of an internal symmetry group $G$ to its subgroup $H$. Let $\lie g$ and $\lie h$ denote the Lie algebras of $G$ and $H$, respectively. Following ref.~\cite{Brauner:2025rzv}, we will assume from the outset that the coset space $G/H$ is symmetric, which requires the existence of an involutive automorphism $\mathcal{R}:\lie g\to\lie g$ such that $\lie h$ is the eigenspace with eigenvalue $+1$. We will use the symbol $\lie m$ for the eigenspace with eigenvalue $-1$. A basis of $\lie g$ will be denoted as $Q_A,Q_B,\dotsc$. It can always be chosen so that it splits into a basis $Q_\a,Q_\b,\dotsc$ of $\lie h$ and a basis $Q_a,Q_b,\dotsc$ of $\lie m$. The presence of the automorphism $\mathcal{R}$ then forces the commutation relations among the generators to take the form
\begin{equation} 
[Q_\a,Q_\b]=\imag f^\g_{\a\b}Q_\g,\qquad
[Q_\a,Q_b]=\imag f^a_{\a b}Q_a,\qquad
[Q_a,Q_b]=\imag f^\a_{a b}Q_\a.
\label{NLSM:commutation} 
\end{equation}
We will further assume $G$ to be semisimple. The Lie algebra $\lie g$ then admits a nondegenerate symmetric bilinear form $g_{AB}$ that is both $\mathcal{R}$-invariant and $G$-invariant. The former implies that the broken and unbroken subspaces $\lie h$ and $\lie m$ are mutually orthogonal, i.e.~$g_{\a a}=0$. Accordingly, upon restriction to the corresponding subspaces, the bilinear form $g_{AB}$ splits into $g_{\a\b}$ and $g_{ab}$. Moreover, the $G$-invariance of $g_{AB}$ allows one to define $f_{ABC} \equiv g_{AD} f^{D}_{BC}$, which is fully antisymmetric.

The elementary fields of the NLSM are the NG bosons $\pi^a$, parametrising the coset space $G/H$. These can be chosen to transform linearly under the unbroken subgroup $H$, provided the latter is compact~\cite{Brauner2024}. It is common to encode the NG variables in a $G$-valued coset representative $U(\pi)$ such that the lift of $\mathcal{R}$ to $G$ maps $U(\pi)$ to $U(\pi)^{-1}$, and for $\pi^a=0$, $U(\pi)$ reduces to unity. Otherwise, the choice of the coset representatives is arbitrary. One can then construct a set of secondary, composite operators, assembled componentwise in the $\lie g$-valued Maurer-Cartan (MC) form,
\begin{equation}
\mathbf{\O}(\pi) \equiv-\imag U(\pi)^{-1}\dd U(\pi)\equiv \mc^a(\pi) Q_a- A^\a(\pi) Q_\a.
\label{MCform}
\end{equation}
The unconventional minus sign in front of the second term will somewhat simplify the duality map between the NLSM and the YMS theory, discussed below. This second part of the MC form transforms as an $H$-connection, and thus allows for the definition of the associated field strength and covariant derivatives,
\begin{equation}
F^\a_{\m\n}\equiv\de_\m A^\a_\n-\de_\n A^\a_\m + f^\a_{\b\g} A^\b_\m A^\g_\n,\qquad
\Da_\m\mc^a_\n\equiv\de_\m\mc^a_\n + f^a_{\a b} A^\a_\m\mc^b_\n.
\label{NLSM:GDmcdef}
\end{equation}
These are subject to the geometric identities
\begin{equation}
F^\a_{\m\n} = - f^\a_{ab}\mc^a_\m\mc^b_\n, \qquad \D_\m\mc^a_\n = \D_\n\mc^a_\m,
\label{NLSM:geomids}
\end{equation}
which descend directly from the MC structure equation, $\dd \mathbf{\O}=-\tfrac{\imag}{2}[\mathbf{\O},\mathbf{\O}]$.

Turning to dynamics, the lowest-order, two-derivative Lagrangian of the NLSM is constrained by the spontaneously broken symmetry to take the form
\begin{equation}
\L_\text{NLSM} = \tfrac 12\k_{ab}\mc^a_\m\mc^{b\m},
\label{NLSM:Lagrangian}
\end{equation}
where $\k_{ab}$ is a constant, symmetric and positive-definite matrix, required to be invariant under the adjoint action of $H$. The resulting primary EoM for the NG fields can be succinctly written as a covariant conservation law for the broken part of the MC form,
\begin{equation}
    \D_\m\mc^{a\m}=0.
    \label{NLSM:conservation}
\end{equation}
By combining this with the geometric constraints~\eqref{NLSM:geomids}, one can derive a pair of secondary field equations that are of second order in $\mc^a_\m$ and $A^\a_\m$, respectively~\cite{Brauner:2025rzv},
\begin{equation}
     \boxedcolor{Green}{\D^2\mc^a_\m +\eta^{\r \s}  f^a_{\a b}f^\a_{cd}\mc^b_\r\mc^{c}_\s\mc^d_\m=0, \qquad 
  \D_\m F^{\a\m\n} - \eta^{\r \s} f^\a_{ab}(\D^\n\mc^a_\r)\mc^{b}_\s= 0.}
  \label{NLSM:eom}
 \end{equation}
It is precisely this covariant formulation of the NLSM that enables a direct map to the YMS theory.


\subsection{Yang-Mills-scalar theory}
\label{sec:YMS}

The YMS theory describes interactions of $H$-valued gauge bosons $A_\m^\a$ with scalar fields $\p^{a}_{I}$~\cite{Cachazo2015}. The latter transform under both the gauge group $H$ in a suitable representation $\mathcal{V}$ (labelled by $a$) and under an additional global symmetry group $G_\text{flavour}$ (labelled by $I$). In anticipation of the duality map to the NLSM, we choose the specific Lagrangian
\begin{equation}
 \L_\text{YMS} = -\tfrac14g_{\a\b}F^\a_{\m\n}F^{\b\m\n}+\tfrac12g^{IJ}g_{ab}\D_\m\p_I^{a}\D^\m\p_J^{b}-\tfrac{1}{4} g^{IJ}g^{KL}g_{\a\b}f^\a_{ac}f^\b_{bd}\p_I^{a}\p_J^{b}\p_K^{c}\p_L^{d},
 \label{YMS:lagrangian}
\end{equation}
where the field strength and the covariant derivative are given by
 \begin{equation}
     F^\a_{\m\n} = \de_\m A^\a_\n - \de_\n A^\a_\m + f^\a_{\b\g} A_\m^\b A_\n^\g, \qquad \Da_\m\p_I^{a}=\de_\m\p_I^{a} + f^a_{\a b}A^\a_\m\p_I^{b}.
 \end{equation}
While $f^\a_{\b\g}$ are naturally the structure constants of $H$, the coefficients $f^a_{\a b}$ should, for the time being, be thought of as a shorthand notation for the matrices that represent the generators of $H$ on the space $\mathcal{V}$ of the scalar fields. Likewise, $f^\a_{ab}$ are, for the time being, merely a set of constants defining uniquely the scalar self-interactions. In order for the theory~\eqref{YMS:lagrangian} to describe propagating excitations above a stable ground state, the various matrix coefficients therein must satisfy a set of positivity constraints. For the kinetic term of the gauge fields to be positive-definite, we assume the gauge group $H$ to be compact and choose $g_{\a\b}$ as positive-definite. Similarly, to have a positive-definite kinetic term for the scalars, we assume the flavour symmetry group $G_{\text{flavour}}$ to be compact. Moreover, the flavour metric $g_{IJ}$ and the bilinear form $g_{ab}$ on $\mathcal{V}$ should both be definite, and their signatures should be correlated,
\begin{equation}
    \gr{sign}(g_{IJ})=\gr{sign}(g_{ab}).
    \label{eq:YMS_sign}
\end{equation}

The geometric duality allows one to relate the (primary) field equations of the YMS theory to the secondary equations~\eqref{NLSM:eom} of the NLSM. To that end, we assume that the ``metrics'' $g_{\a\b}$ and $g_{ab}$ in the YMS theory can be used to raise and lower indices on $f^\a_{\b\g}$, $f^\a_{ab}$ and $f^a_{\a b}$, and that $f$s with three subscripts are fully antisymmetric. This can be ensured, for instance, by treating the collection $f^\a_{\b\g}$, $f^\a_{ab}$, $f^a_{\a b}$ in the YMS theory as actual counterparts of the semisimple Lie algebra $\lie g$ in the NLSM, and $g_{\a\b}$, $g_{ab}$ as the components of the corresponding $G$-invariant bilinear form $g_{AB}$ on $\lie g$. With the above qualifications, the EoMs of the YMS theory can be expressed as
\begin{equation}
 \boxedcolor{Red}{\D^2\p^{a}_I + g^{JK}f^a_{\a b}f^\a_{cd}\p_J^{b}\p_K^{c}\p_I^{d} = 0, \qquad 
 \D_\m F^{\a\m\n} - g^{IJ} f^\a_{a b}(\D^\n\p_I^{a})\p_J^{b} = 0.}
 \label{YMS:eom}
 \end{equation}
One now immediately recognises the structural similarity between eqs.~\eqref{NLSM:eom} and~\eqref{YMS:eom}. Indeed, the two sets of equations are related by the following map,
\begin{equation}
     \text{NLSM:}\,\,\{\mc^a_\m, A^\a_\m\} \longrightarrow \text{YMS:}\,\,\{\p_I^{a}, A^\a_\m\}.
     \label{map-NLSM-YMS}
 \end{equation}
This map underlies the geometric duality between the two theories. 

An important point is that, to render this map consistent, one has to identify the spacetime index $\m$ carried by the broken MC components $\mc^a_\m$ with the flavour index $I$ of the YMS scalars $\p^a_I$. Such an identification is meaningful only if the corresponding metrics are likewise mapped to each other. This appears problematic at first sight, since the spacetime metric $\eta_{\m\n}$ is indefinite whereas the flavour metric $g_{IJ}$ is required to be definite. A natural resolution is to restrict to solutions on the NLSM side with purely spatial dependence, such that all the MC components carry spatial indices only. Decomposing $\m=(0,i)$, we thus identify the flavour index $I$ with the spatial index $i$ and set
\begin{equation}
g_{IJ}=\d_I^i\d_J^j\eta_{ij}=-\d_{IJ}.
\label{eq:YMS_flavour metric map}
\end{equation}
The scalar flavours are then in a one-to-one correspondence with spatial dimensions, $I,J\in\{1,\dots,D-1\}$. In this way, static solutions of the NLSM will be mapped to static solutions of the YMS theory with an~$\gr{SO}(D-1)$ flavour symmetry.\footnote{There is also a temporal branch of the duality map, obtained by restricting to a single scalar flavour and taking $g_{IJ}=\eta_{00}=+1$. This maps uniform (space-independent) solutions of the NLSM to uniform solutions of the YMS theory with one scalar flavour. We do not pursue this trivial possibility any further.} Alternatively, one can say that \emph{any} solution of the NLSM in Euclidean space maps to a corresponding solution of the Euclidean YMS theory. Interestingly, the condition~\eqref{eq:YMS_sign} implies $\operatorname{sign}(g_{ab})=\operatorname{sign}(g_{IJ})=-1$. Since we already assumed $g_{\a\b}$ to be positive-definite, consistency of the duality map therefore requires the bilinear forms $g_{\a\b}$ and $g_{ab}$ to have opposite signatures. In order to land on a physical YMS theory with a stable vacuum, we accordingly have to start with a NLSM based on a noncompact coset space $G/H$, for compact $G$ would require the whole $g_{AB}$ to be definite.

Another aspect of the duality map is that on the NLSM side, the MC components $\mc^a_\m$ and $A^\a_\m$ descend from the elementary NG fields $\pi^a$. Under the geometric duality~\eqref{map-NLSM-YMS}, these composite operators are mapped to the elementary fields of the YMS theory. A nontrivial remnant of this distinction lies in the geometric identities~\eqref{NLSM:geomids} together with the covariant conservation law~\eqref{NLSM:conservation}, which translate into the constraints
\begin{equation}
  F^\a_{ij} = - \d_i^I \d_j^J f^\a_{ab}\p^a_I \p^b_J , \qquad \d_j^J D_i \p^a_J=  \d_i^I D_j \p^a_I, \qquad \d^{i J} D_i \p^a_J=0
 \label{YMS-condition}
 \end{equation}
on the fields on the YMS side. Consequently, the geometric duality maps (static) solutions of the NLSM to a subset of (static) solutions of the YMS theory. Configurations of the YMS fields not satisfying the relations~\eqref{YMS-condition} cannot be recovered from the NLSM.

To quickly summarize, geometric duality allows one to map exact classical solutions of the NLSM to exact classical solutions of the YMS theory. Starting from the NLSM~\eqref{NLSM:Lagrangian} on the coset space $G/H$ with compact unbroken subgroup $H$ but noncompact full symmetry group $G$, and restricting to static solutions depending on the $D-1$ spatial coordinates, one ends up with static solutions of the YMS theory~\eqref{YMS:lagrangian} with gauge group $H$, a particular quartic scalar self-interaction, and an additional $\gr{SO}(D-1)$ flavour symmetry.


\section{Classical solutions under NLSM-YMS duality}
\label{sec:Sol_Map1}

In this section, we illustrate the duality between the NLSM and the YMS theory by working out a class of sample solutions of the NLSM and their images under the duality map. To that end, we make use of known exact solutions of the NLSM in two (spatial) dimensions. We review these solutions first, and then, in a follow-up subsection, detail the corresponding solutions of the YMS theory. 


\subsection{Exact solutions of NLSM in two dimensions}

One of the most well-studied examples of NLSM is that on the two-dimensional sphere, $S^2\simeq\gr{SU}(2)/\gr{U}(1)$, describing physically spin waves in antiferromagnets. However, the coset space of this NLSM is compact, which poses an obstacle to mapping the solutions to a physical YMS theory. To circumvent this problem, we can switch to the noncompact cousin of the sphere: the hyperboloid $H^2\simeq\gr{SU}(1,1)/\gr{U}(1)$. The NLSM on the latter is sometimes referred to as the Cremmer-Julia NLSM~\cite{Cremmer:1979up}. It will prove instructive to consider the two NLSMs and their solutions alongside each other. We will restrict ourselves to three spacetime dimensions, and will work out an infinite class of exact static solutions of the two NLSMs.

Starting with the Lie algebra $\lie g$, we will use the basis of generators $Q_A\equiv\t_A/2$ for $\gr{SU}(2)$. On the $\gr{SU}(1,1)$ side, we will use the generators $Q_{1,2}\equiv\imag\t_{1,2}/2$ and $Q_3\equiv\t_3/2$. In both cases, $Q_3$ is the generator of the unbroken $\gr{U}(1)$ subgroup. The coset spaces are symmetric with the involutive automorphism $\mathcal{R}$ given by conjugation with $\t_3$. The structure of both Lie algebras can be encoded jointly in the commutation relations
\begin{equation}
[Q_3,Q_a]=\imag\varepsilon_a^{\phantom ab}Q_b,\qquad
[Q_a,Q_b]=\imag\s\varepsilon_{ab}Q_3,
\end{equation}
where $a,b\in\{1,2\}$, $\epsilon_{12}=1$, with $\s=+1$ for the compact case $\gr{SU}(2)$ and $\s=-1$ for the noncompact case $\gr{SU}(1,1)$. In both cases, the symmetry group $G$ is simple, and thus admits a unique $G$-invariant bilinear form $g_{AB}$, equal to $\tr(Q_AQ_B)$ up to a choice of normalization. 

Both coset spaces are two-dimensional and can be parametrised by a complex variable $\psi$, charged under the unbroken $\gr{U}(1)$ symmetry. We adopt the following stereographic parametrisation of the two coset spaces,
\begin{equation}
U(\psi)\equiv\frac1{\sqrt{1+\s\psi\bpsi}}[\un+(\psi-\bpsi)Q_1-\imag(\psi+\bpsi)Q_2],
\label{Upsi}
\end{equation}
similar to a parametrisation used in ref.~\cite{Deger:1999st}. The individual components of the MC form, upon expansion in the basis of generators $Q_A$, are
\begin{equation}
\mc^1=\frac1\imag\frac{\dd\psi-\dd\bpsi}{1+\s\psi\bpsi},\qquad
\mc^2=-\frac{\dd\psi+\dd\bpsi}{1+\s\psi\bpsi},\qquad
A = \imag\s\frac{\bpsi\dd\psi-\psi\dd\bpsi}{1+\s\psi\bpsi}.
\label{MCformcomp}
\end{equation}
The two-derivative Lagrangian~\eqref{NLSM:Lagrangian} is determined uniquely up to overall normalization,
\begin{equation}
\L_\text{NLSM} \propto \tfrac14\d_{ab}\mc^a_\m\mc^{b\m} = \frac{ \de_\m\psi \de^\m\bpsi}{(1+\s\psi\bpsi)^2}.
\label{S2metric}
\end{equation}

There is a considerable amount of literature on various exact solutions of our two NLSMs, especially the compact version, $\gr{SU}(2)/\gr{U}(1)$. The brief overview below is closest in spirit to refs.~\cite{Gross:1977wu,Tseitlin:1983,Ody:1994pk}; a particularly clear presentation of the compact case can be found in ref.~\cite{Pastras:2016yaa}. Restricting ourselves to static solutions, the spatial dependence of the solutions can be packaged into (anti)holomorphic coordinates $\xi \equiv x^1 + \imag x^2$ and $\bar \xi \equiv x^1 - \imag x^2$. The field equations for static solutions of the NLSMs then take the form\footnote{The $\s=-1$ version of this EoM is known as the Ernst equation, which describes exact axially symmetric solutions to Einstein equations in general relativity~\cite{Ernst:1967wx}, as well as some solutions of the Einstein-Maxwell theory~\cite{Ernst:1967by}.}
\begin{equation}
(1+\s\psi\bpsi)\de_\xi\de_{\bxi}\psi-2\s\bpsi\de_\xi\psi\de_{\bxi}\psi=0.
\label{NLSM:complexEoM}
\end{equation}
An immediate consequence is that any holomorphic ($\de_{\bxi}\psi=0$) or antiholomorphic ($\de_\xi\psi=0$) function trivially solves the EoM. These are the well-known skyrmion and antiskyrmion solutions, which exist for both compact and noncompact cases. Beyond such purely (anti)holomorphic solutions, there is a much larger family with an interesting interpretation. To that end, we introduce the more general ansatz
\begin{equation}
\psi(\xi,\bxi)=\c(\xi)f(\abs{\c(\xi)}^2),
\label{meron:ansatz}
\end{equation}
where $\c$ is a holomorphic function and $f$ is a real function of a real variable. Note that our ansatz encompasses both skyrmion solutions, recovered by setting $f(x)=1$, and antiskyrmion solutions, recovered from $f(x)=1/x$. It reduces the EoM~\eqref{NLSM:complexEoM} to a second-order ordinary differential equation for the profile function $f$,
\begin{equation}
x[1+\s xf^2(x)]f''(x)+2f'(x)[1-\s x^2f(x)f'(x)]=0,
\label{meron:EoM1}
\end{equation}
independently of the choice of $\c$.

In order to further simplify the problem of finding exact solutions of the NLSMs, we need to treat the compact and noncompact cases separately. In case of $S^2\simeq\gr{SU}(2)/\gr{U}(1)$, we employ the redefinition
 \begin{equation}
     x=e^{2y}, \quad f(x)= e^{-y} {\tan\frac{h(y)}4} \quad \Rightarrow \quad h''(y)=\sin h(y)\qquad(\s=+1).
 \end{equation}
The resulting EoM for $h(y)$ is nothing but the sine-Gordon equation, describing the mathematics of a pendulum. It can be solved in a closed form in terms of Jacobi elliptic functions. Simple explicit solutions include the (anti)kink solutions $h(y)=4\arctan e^{\pm y}$, which correspond to the aforementioned (anti)skyrmion solutions of the NLSM. Another simple solution is $h(y)=\pi$, corresponding to $f(x) = 1/\sqrt{x}$ and in turn to $\psi(\xi,\bxi)=\sqrt{\c(\xi)/\overline{\c(\xi)}}$. This class of solutions of the NLSM is known as meron.

The noncompact case $H^2\simeq\gr{SU}(1,1)/\gr{U}(1)$ can be treated similarly, except that one now needs two separate substitutions to cover the whole range of possible values of $\psi(\xi,\bxi)$,\footnote{The hyperboloid $H^2$ has two connected components, which in the stereographic coordinates correspond respectively to $|\psi|<1$ and $|\psi|>1$. Solutions taking values from the two components correspond precisely to the powers $\pm1$ in the substitution for $f(x)$.}
\begin{equation}
 x=e^{2y}, \quad
 f(x)= e^{-y} \left[\tanh\frac{{h(y)}}{4}\right]^{\pm 1}  \quad \Rightarrow \quad h''(y)=\sinh h(y)\qquad(\s=-1).
\end{equation}
This is the sinh-Gordon equation. It has exact solutions $h(y)=4\operatorname{arctanh}e^{\pm y}$ similar to the sine-Gordon (anti)kink, which again map to (anti)skyrmion solutions of the NLSM. However, in the present case, there is no analogous of meron solutions. If desired, other solutions can be expressed directly in terms of Jacobi elliptic functions.

We conclude that both versions of the NLSM feature a large family of exact static solutions $\psi(\xi,\bxi)$, parametrised by an arbitrary holomorphic function $\c(\xi)$ and a real function $f(\abs{\c}^2)$ that can be related to the sine- and sinh-Gordon models. Given any solution for the latter, one chooses an arbitrary $\c(\xi)$ and then reconstructs $\psi(\xi, \bxi)$ from eq.~\eqref{meron:ansatz}, which in turn leads to the geometric data of the NLSM via eq.~\eqref{MCformcomp}. 


\subsection{Corresponding solutions of Yang-Mills-scalar theory}

The geometric data of the NLSM are mapped by the duality to the elementary scalar and gauge fields of the YMS theory. In case of the two specific NLSMs discussed above, the YMS theory will have a $\gr{U}(1)$ gauge group and two flavours of coloured scalars $\p_I^a$, where $a$ indicates the two-dimensional real representation of $\gr{U}(1)\simeq\gr{SO}(2)$. We will trade this vector representation for the equivalent one-dimensional complex representation. The correspondence requires that $\p^a_I$ be combined into complex fields, $\Phi_I\equiv\p^1_I+\imag\p^2_I$. In addition to the $\gr{U}(1)$ gauge invariance, the YMS theory has an $\gr{SO}(2)$ flavour symmetry, acting on the flavour indices $I$, inherited from the spatial symmetry of the NLSM in three spacetime dimensions. Below, we will often collect the two complex scalar fields $\Phi_I$ into a complex vector, $\vec\Phi$.

Upon matching carefully the structure constants of the symmetry group and the metrics $g_{\a\b}$, $g_{ab}$ and $g_{IJ}$, we find that the general YMS Lagrangian~\eqref{YMS:lagrangian} boils down to
\begin{equation}
\L_\text{YMS}=-\frac14F_{\m\n}F^{\m\n}-\frac\s2\D_\m\vec\Phi^*\cdot\D^\m\vec\Phi-\frac18\bigl[(\vec\Phi^*\cdot\vec\Phi)^2-\abs{\vec\Phi\cdot\vec\Phi}^2\bigr],
\label{YMSU(1):Lagrangian}
\end{equation}
where the dot product is defined using the positive-definite Euclidean metric, $\vec\Phi^{*}\!\cdot\!\vec\Phi \equiv \d^{IJ}\Phi_I^*\Phi_J$ and analogously for $\vec\Phi\cdot\vec\Phi$. The covariant derivative of the complex scalar $\vec\Phi$ is defined as usual by $\D_\m\vec\Phi\equiv(\de_\m+\imag A_\m)\vec\Phi$ and $\D_\m\vec\Phi^{*}\equiv(\de_\m-\imag A_\m)\vec\Phi^{*}$. The presence of the sign $\s$ in the Lagrangian~\eqref{YMSU(1):Lagrangian} makes it clear that it is the noncompact version of the NLSM with $\s=-1$ that maps to the physical YMS theory with a stable ground state, whereas the compact NLSM with $\s=1$ maps to a theory with a negative-definite kinetic term for the scalars. However, at the level of the EoMs, which in the present case read
\begin{equation}
\D^2\vec\Phi-\frac\s2\bigl[(\vec\Phi^*\cdot\vec\Phi)\vec\Phi-(\vec\Phi\cdot\vec\Phi)\vec\Phi^*\bigr]=\vec0,\quad
\de_\n F^{\m\n}+\frac{\imag\s}2\bigl(\D^\m\vec\Phi^*\cdot\vec\Phi-\vec\Phi^*\cdot\D^\m\vec\Phi\bigr)=0,
\label{YMS:U(1)EoM}
\end{equation}
exact (static) solutions of the NLSM map to exact solutions of the YMS theory for both choices of $\s$.

Combining the general map~\eqref{map-NLSM-YMS} with the expressions~\eqref{MCformcomp} for the components of the MC form, we obtain the elementary fields of the YMS theory in terms of the complex field $\psi$ of the NLSM,
\begin{equation}
\Phi_I=\frac{-2\imag\d_I^i\de_i\psi}{1+\s\psi\bpsi},\qquad
A_i=\imag\s\frac{\bpsi\de_i\psi-\psi\de_i\bpsi}{1+\s\psi\bpsi}.
\end{equation}
It is straightforward if somewhat tedious to verify that these indeed constitute a static solution of the YMS field equations~\eqref{YMS:U(1)EoM} provided the function $\psi(\xi,\bxi)$ satisfies the NLSM EoM~\eqref{NLSM:complexEoM}. Therefore, all solutions of the latter map to solutions of the former.

One can get a somewhat better intuitive insight into the nature of these solutions by computing a set of composite operators that play an important role in the Lagrangian and EoMs: the flavor-singlets $\vec\Phi^*\cdot\vec\Phi$ and $\vec\Phi\cdot\vec\Phi$, and the sole independent component of the field-strength tensor, $F_{xy}$. We list the values of these operators for the specific solutions highlighted previously:
\begin{itemize}
\item For skyrmion solutions on the NLSM side, $\psi(\xi,\bxi)=\c(\xi)$, we find
\begin{equation}
\vec\Phi^*\cdot\vec\Phi=\frac{8\abs{\c'}^2}{(1+\s\abs\c^2)^2},\qquad
\vec\Phi\cdot\vec\Phi=0,\qquad
F_{xy}=-\frac{4\s\abs{\c'}^2}{(1+\s\abs\c^2)^2}.
\end{equation}
\item Antiskyrmion solutions of the NLSM can be represented by $\psi(\xi,\bxi)=\overline{\varphi(\xi)}$ with holomorphic $\varphi$, resulting in 
\begin{equation}
\vec\Phi^*\cdot\vec\Phi=\frac{8\abs{\varphi'}^2}{(1+\s\abs\varphi^2)^2},\qquad
\vec\Phi\cdot\vec\Phi=0,\qquad
F_{xy}=\frac{4\s\abs{\varphi'}^2}{(1+\s\abs\varphi^2)^2}.
\end{equation}
\item Finally, meron solutions of the compact ($\s=1$) version of the NLSM correspond to $\psi(\xi,\bxi)=\sqrt{\c(\xi)/\overline{\c(\xi)}}$, which leads to
\begin{equation}
\vec\Phi^*\cdot\vec\Phi=\frac{\abs{\c'}^2}{\abs\c^2},\qquad
\vec\Phi\cdot\vec\Phi=\frac{\abs{\c'}^2}{\bar\c^{2}},\qquad
F_{xy}=0.
\end{equation}
While the field-strength tensor on the YMS side vanishes, the gauge field still has a nontrivial structure. To see this, take the gauge potential,
\begin{equation}
A_x=\frac\imag2\left(\frac{\c'}{\c}-\frac{\bar\c'}{\bar\c}\right),\qquad
A_y=-\frac12\left(\frac{\c'}{\c}+\frac{\bar\c'}{\bar\c}\right),
\end{equation}
integrate over a closed loop in the coordinate space, and convert the integral into one in the complex plane,
\begin{equation}
\oint\vek A\cdot\dd\vek x=\frac\imag2\oint\left(\frac{\c'}{\c}\dd \xi-\frac{\bar\c'}{\bar\c}\dd \bxi\right)=-\operatorname{Im}\oint\frac{\c'}{\c}\dd \xi.
\end{equation}
This integral receives a contribution $-2\pi$ from every zero of $\c(\xi)$ and $+2\pi$ from every pole of $\c(\xi)$. The gauge field descending from meron solutions of the NLSM therefore describes a combination of point vortices and antivortices, located respectively at the zeros and poles of $\c(\xi)$.
\end{itemize}


\section{Equations of motion for the sGal-DBI-NLSM\textsubscript{g} duality sequence}
\label{sec:EoMoverview2}

In this section, we follow the same logical strategy as above and formulate the geometric duality connecting the sGal and DBI theories and the NLSM coupled to dynamical gravity. We will use systematically the method of nonlinear realizations for spontaneously broken symmetry aka coset construction. This will allow us to construct the geometric data of the three theories from the MC form, and subsequently place them in direct correspondence, enabling us to map exact solutions in the same spirit as we did for the NLSM and the YMS theory. In particular, the composite metrics of the sGal and DBI theories thus become genuinely propagating gravitational fields in the NLSM\textsubscript{g}. In the final subsection, we develop a complementary approach, exploiting a judicious choice of the coset representatives, whereby the fundamental scalar fields of the sGal-DBI-NLSM\textsubscript{g} duality sequence can be mapped to one another directly following a remarkably simple derivative hierarchy.


\subsection{Special Galileon theory}
\label{sec:sGal}

We begin with the sGal theory~\cite{Cheung2015a,Hinterbichler2015a}, playing the role of the seed theory in the sGal-DBI-NLSM\textsubscript{g} duality sequence. The elementary degree of freedom of the sGal theory is a single scalar field $\p$. However, we will need the geometric data of the theory, conveniently supplied by the coset construction. Below, we review briefly the pertinent parts of the construction, focusing on the geometric aspects insensitive to the choice of field parametrisation. For full details, see section 5 of the prequel~\cite{Brauner:2025rzv} to this paper.

The symmetry algebra of the sGal theory includes the Poincar\'e generators $L_{\m\n}$ and $P_\m$, a scalar generator $Q$, a vector $K_\m$, and a traceless symmetric tensor $S_{\m\n}$. The latter satisfies the commutation relation
\begin{equation}
  [S_{\m\n},S_{\k\l}]=\imag\s(\eta_{\m\l}L_{\n\k}+\eta_{\n\k}L_{\m\l}+\eta_{\m\k}L_{\n\l}+\eta_{\n\l}L_{\m\k}),
\end{equation}
where the sign $\s=\pm1$ distinguishes between two versions of the sGal theory, corresponding respectively to a nonlinearly realised $\gr{SL}(D,\R)$ for $\s=+1$~\cite{Roest2021}, and nonlinearly realised $\gr{SU}(D-1,1)$ for $\s=-1$~\cite{Novotny2017a}. All other commutators of the symmetry algebra are fixed by consistency constraints imposed by the Jacobi identity~\cite{Bogers2018b}.

In the coset construction, a set of NG variables must be introduced for all the nonlinearly realized symmetries. In case of the sGal theory, this includes the spacetime translations $P_\m$ (where the corresponding coset variables are simply the spacetime coordinates $x^\m$), the scalar $Q$, the vector $K_\m$, and the traceless symmetric tensor $S_{\m\n}$. Upon a largely arbitrary choice of the coset representative, the MC form is introduced along the same lines as in the NLSM in section~\ref{sec:NLSM}, and decomposed in the basis of the Lie algebra of the symmetry group,
\begin{equation}
\mathbf{\O}\equiv \tfrac12\tilde{\O}^{\a\b}S_{\a\b}+\mc^\a_KK_\a+\mc_QQ+\mc^\a_PP_\a-\tfrac12 \mc^{\a\b}_LL_{\a\b},
\label{eq:sGal_MC}
\end{equation}
where the minus sign in front of the last term is purely conventional. The NG mode for the scalar generator $Q$ is identified with the elementary scalar field $\p$ of the sGal theory. The NG fields for $K_\m$ and $S_{\m\n}$ are mere auxiliary degrees of freedom, and are eventually eliminated by imposing a set of inverse Higgs constraints~\cite{GarciaSaenz2019},
\begin{equation}
    \begin{split}
        \mc_Q=0,\qquad
    \mc_K^\a = \X \mc^\a_P,
    \end{split}
    \label{eq:sGal_IHCs}
\end{equation}
where $\X$ is a geometric scalar that will later turn out to play a key role in the EoM of the sGal theory. The remaining independent components of the MC form constitute the geometric data of the sGal theory.

The component $\mc^{\a\b}_L$ establishes a spin connection that can be used to define derivatives of tensor fields, covariant under the Lorentz group. The component $\mc^\a_P$ defines via $\mc^\a_P\equiv e^\a_\m\dd x^\m$ a coframe, $e^\a_\m$. With respect to this coframe, however, the spin connection has a nonvanishing torsion. This issue can be remedied by rescaling with a suitable scalar factor,
\begin{equation}
\E^\a_\m\equiv e^\a_\m\sqrt{1-\s \X^2}.
\end{equation}
Using the improved coframe $\E^\a_\m$, one can build an effective metric,
\begin{equation}
G_{\m\n}\equiv\E^{\a}_\m\eta_{\a\b}\E^{\b}_\n,
\end{equation}
whose Levi-Civita connection corresponds to the spin connection $\mc^{\a\b}_L$. Finally, the component $\tilde\O^{\a\b}$ of the MC form can be combined with the scalar $\X$ to define a traceful symmetric-tensor one-form,
\begin{equation}
    \O^{\a\b}\equiv \tilde{\O}^{\a\b}-\frac{\eta^{\a\b}}{1-\s\X^2}\dd\X,\label{eq:sGal_tensor_def}
\end{equation}
which, upon pull-back to the spacetime, defines a fully symmetric rank-3 tensor $\O_{\m\n\l}\equiv\E_{\a\m}\E_{\b\n}\O^{\a\b}_{\l}$. This serves as a suitable interpolating field for the sGal scalar $\p$, being at the linear order proportional to $\de_\m\de_\n\de_\l\p$. The metric $G_{\m\n}$ and the tensor $\O_{\m\n\l}$ together represent the primary geometric data of the sGal theory. For the purposes of establishing the geometric duality between the sGal theory, the DBI theory and the NLSM\textsubscript{g}, it will however be convenient to keep in mind the tensor $\O^{\a\b}_\m$ with two frame indices. Converting spacetime indices to frame indices requires a choice of frame, and thus implies an inherent gauge redundancy, similar to the gauge redundancy in the NLSM.

The MC structure equations imply a set of geometric identities for the geometric data of the sGal theory, analogous to eq.~\eqref{NLSM:geomids},
\begin{align}
    \label{eq:sGal_Gauss}
    &\boxedcolor{Green}{R^{\a \b}{}_{\m \n} =-\s \eta_{\a'\b'} \bigl( \O^{\a\a'}_\m \O^{\b\b'}_\n - \O^{\a\a'}_\n \O^{\b\b'}_\m \bigr)}&&\text{(Gauss)},\\
    \label{eq:sGal_Codazzi}
    &\boxedcolor{Green}{\DD_\m\O^{\a\a'}_{\n}=\DD_\n\O^{\a\a'}_{\m}}&&\text{(Codazzi)},
\end{align}
where the covariant derivative acts as
\begin{equation}
\DD_\m\O^{\a\a'}_{\n}=\nabla_\m\O^{\a\a'}_{\n}+\eta_{\b\g}\mc^{\a\b}_{L\m}\O^{\g\a'}_{\n}+\eta_{\b'\g'}\mc^{\a'\b'}_{L\m}\O^{\a\g'}_{\n}.\label{sGal_covD_def}
\end{equation}
At this stage, primed and unprimed Greek indices are treated on equal footing. However, in view of the duality map to the DBI theory detailed below, we point out that the sGal tensor $\O^{\a\a'}_\m$ can always be written with one unprimed and one primed index. The composite spin connection, implicit in the covariant derivative, then carries either two unprimed indices or two primed ones.

The Gauss relation~\eqref{eq:sGal_Gauss} implies that the curvature of the effective metric is fully determined by $\O^{\a\b}_{\m}$, so that the would-be degrees of freedom contained in $G_{\m\n}$ are not independent, but rather composite operators related to the elementary scalar $\p$. The primary sGal field equation for $\p$ turns out to be simply $\X=0$. Thus, on solutions of the EoM, the tensor $\O_{\m\n\l}$ is traceless, or equivalently
\begin{equation}
\boxedcolor{Green}{\O^{\a}\equiv\E_{\b}^\m\O^{\a\b}_{\m}=0.}
\label{sGal:EoMtrace}
\end{equation}
This serves as the secondary covariant EoM for the sGal theory. By taking additional (covariant) derivatives and using the geometric Gauss and Codazzi identities, it is possible to derive a set of further, second-order equations, one for the covariant ``matter'' field $\O^{\a\b}_\m$, one for the spin connection, and finally an Einstein-type equation for the composite metric, whose right-hand side depends solely on the matter tensor $\O^{\a\b}_\m$. However, we choose not to list these equations explicitly for we will not need them to map the exact classical solutions. To that end, we will only need the identities~\eqref{eq:sGal_Gauss}--\eqref{eq:sGal_Codazzi} and the secondary EoM~\eqref{sGal:EoMtrace}.


\subsection{Dirac-Born-Infeld theory}
\label{sec:DBI}

We now turn to the DBI theory with $n$ flavours of scalars $\p^I$, $I=1,\dotsc,n$, which describes a $D$-dimensional brane fluctuating in an ambient flat $(D+n)$-dimensional spacetime. As in the sGal case, a fully covariant formulation of the DBI theory can be obtained via the coset construction. We again briefly review the pertinent geometric aspects of the construction, and refer the reader to ref.~\cite{Brauner:2025rzv} for further details.

The starting point is the collection of all the symmetry generators. Besides the Poincar\'e generators $L_{\m\n}$ and $P_\m$, these include two sets of Lorentz scalars: the (broken) translations $Q_I$ in the extra dimensions, and the (unbroken) extra-dimensional rotations $Q_{IJ}$, spanning a flavour group $\gr{SO}(n)$. Finally, there is a collection of Lorentz vectors $K_{\m I}$, corresponding geometrically to rotations that connect the physical and extra dimensions. Altogether, the symmetry spanned on all the generators is either $\gr{ISO}(D+n-1,1)$ or $\gr{ISO}(D-1,n+1)$, depending on the signature of the metric $g_{IJ}$ in the extra dimensions, which we restrict to be definite and hence (after a suitable change of basis) $g_{IJ}=\pm \d_{IJ}$.\footnote{The version of the DBI theory with extra timelike dimensions is not consistent with positivity bounds, required for the existence of an ultraviolet completion~\cite{Adams:2006sv}. However, here we will treat both versions of the DBI theory agnostically on equal footing.} A positive-definite choice corresponds to extra spacelike dimensions, whereas a negative-definite choice corresponds to extra timelike dimensions. The full symmetry group is spontaneously broken down to $\gr{ISO}(D-1,1)\times\gr{SO}(n)$, that is, a direct product of the  Poincar\'e group and a group of flavour rotations.

The nonlinearly realized symmetries that determine the relevant coset space of the DBI theory include all the symmetries that are actually spontaneously broken (those generated by $Q_I$ and $K_{\m I}$) plus spacetime translations. Upon a choice of coset representative, which includes geometrically the choice of embedding of the brane in the ambient space, the MC form is constructed as usual, and decomposed in the basis of generators,
\begin{equation}
\mathbf{\O}\equiv\K^{\a I}K_{\a I}+\mc^I_QQ_I+\mc^\a_PP_\a-\tfrac12\mc^{\a\b}_L L_{\a\b}-\tfrac12 A^{IJ}Q_{IJ}.
\end{equation}
Similarly to the sGal case, the NG modes for the scalar generators are identified with the elementary scalar fields $\p^I$ of the DBI theory. The NG fields for $K_{\a I}$ are auxiliary, and can be eliminated by imposing the inverse Higgs constraint $\mc^I_Q=0$. The remaining components of the MC form constitute the geometric data of the DBI theory.

These data include the spin connection $\mc^{\a\b}_L$ and an additional flavour connection $A^{IJ}$ of the unbroken flavour symmetry. Furthermore, there is $\mc^\a_P$, which establishes a coframe $\E^\a_\m$ and in turn an effective metric on the brane via 
\begin{equation}
\mc^\a_P\equiv \E^\a_\m\dd x^\m,\qquad 
G_{\m\n}=\E^\a_\m \eta_{\a\b} \E^\b_\n.
\end{equation}
Finally, the component $\K^{\a I}$ of the MC form defines a rank-2 symmetric Lorentz tensor $\K^I_{\m\n}\equiv\eta_{\a\b}\E^\a_\m\K^{\b I}_{\n}$, which has the interpretation as the extrinsic curvature of the brane~\cite{Brauner:2025rzv,Hinterbichler:2010xn}. This serves as an interpolating field for the DBI scalars $\p^I$, being at the linear order proportional to $\de_\m\de_\n\p^I$. The primary geometric data of the DBI theory, characterizing the local geometry of the brane, are the metric $G_{\m\n}$ and the extrinsic curvature $\K^I_{\m\n}$. However, for the purposes of mapping the DBI theory to the NLSM\textsubscript{g}, it will again be convenient to work with the tensor $\K^{\a I}_\m$ where one of the Lorentz indices is converted to a frame index. This implies an implicit gauge redundancy due to the choice of frame on the brane.

With the help of the MC structure equations, the following geometric identities --- the Gauss, Codazzi and Ricci relations --- can then be established,
\begin{align}
\label{eq:DBI_Gauss}
&\boxedcolor{Green}{\boxedcolor{Red}{R^{\a \b}{}_{\m \n}=-g_{IJ} \bigl( \K^{\a I}_\m \K^{\b J}_\n - \K^{\a I}_\n \K^{\b J}_\m \bigr)}}&&\text{(Gauss)},\\
\label{eq:DBI_Codazzi}
&\boxedcolor{Green}{\boxedcolor{Red}{\DD_\m\K^{\a I}_{\n}=\DD_\n\K^{\a I}_{\m}}}&&\gr{(Codazzi)},\\
\label{eq:DBI_Ricci}
&\boxedcolor{Green}{\boxedcolor{Red}{F^{IJ}_{\phantom{IJ}\m\n}=\eta_{\a\b}(\K^{\a I}_{\m}\K^{\b J}_{\n}-\K^{\a I}_{\n}\K^{\b J}_{\m})}}&&\gr{(Ricci)},
\end{align}
where $F^{IJ}_{\phantom{IJ}\m\n}$ denotes the field strength of the composite flavour connection $A^{IJ}_\m$,
\begin{equation}
F^{IJ}_{\phantom{IJ}\m\n}=\de_\m A^{IJ}_\n-\de_\n A^{IJ}_\m+g_{KL}(A^{IK}_\m A^{JL}_\n-A^{IK}_\n A^{JL}_\m).
\end{equation}
Besides, the covariant derivative in eq.~\eqref{eq:DBI_Codazzi} is defined as
\begin{equation}
\DD_\m\K^{\a I}_{\n}\equiv\nabla_\m\K^{\a I}_{\n}+\eta_{\b\g}\mc^{\a\b}_{L\m}\K^{\g I}_{\n}-g_{JK}A^{IJ}_{\m}\K^{\a K}_{\n}.
\label{DBI:covder}
\end{equation}
The doubly coloured boxes indicate that all three of the above relations will play a role in connecting the DBI theory via duality to both the sGal theory and the NLSM\textsubscript{g}.

The action of the DBI theory is given, up to a choice of overall scale, by integration of the induced volume form on the brane,
\begin{equation}
S_\text{DBI}\propto\int\dd^D\!x\,\sqrt{\abs{\det G}}.
\label{DBI:action}
\end{equation}
The corresponding primary DBI field equation is
\begin{equation}
\boxedcolor{Red}{\K^{I}\equiv \E^{\m}_\a\K^{\a I}_{\m}=G^{\m\n}\K^I_{\m\n}=0,}
\label{eq:DBI_minimal surface}
\end{equation}
which is nothing but the geometric condition that the brane locally minimises or maximises the induced volume defined by the action. Using judiciously the geometric identities~\eqref{eq:DBI_Gauss}--\eqref{eq:DBI_Ricci}, it then takes little effort to obtain the following descendant covariant EoMs for the DBI theory,
\begin{equation}
\begin{aligned}
&\boxedcolor{Green}{\DD^\m\K^{\a I}_{\m}=0}&&\text{(DBI matter)},\\
&\boxedcolor{Green}{R_{\m\n}=g_{IJ}\eta_{\a\b}\K^{\a I}_{\m}\K^{\b J}_{\n}}\qquad&&\text{(Einstein)}.
\end{aligned}
\label{DBI:EoMscalar}
\end{equation}
The first line furnishes a secondary, covariant conservation law for the extrinsic curvature, whereas the second line is an Einstein-type relation
for the composite metric. Similarly to the sGal theory, there are additional relations one can derive by taking further covariant derivatives, notably a second-order descendant EoM for the composite gauge boson of the flavour connection $A^{IJ}_\m$. However, these will not be needed in the following, and we therefore choose not to list them explicitly.

It is now possible to map the geometric identities~\eqref{eq:sGal_Gauss}--\eqref{eq:sGal_Codazzi} of the sGal theory to the corresponding identities~\eqref{eq:DBI_Gauss}--\eqref{eq:DBI_Ricci} in the DBI theory, and the secondary EoM~\eqref{sGal:EoMtrace} to the primary EoM~\eqref{eq:DBI_minimal surface} of the DBI theory. For the correspondence to be precise, we need to make the following replacements,
\begin{align}
     \text{sGal:}\,\,\{\O^{\a\a'}_\m,\mc^{\a'\b'}_{L\m},\mc^{\a\b}_{L\m},\E_\m^\a\} \longrightarrow \text{DBI:}\,\,\{\K^{\a I}_\m, -\s A^{IJ}_{\m},\mc^{\a\b}_{L\m},\E_\m^\a\} .
     \label{map-sGal-DBI}
 \end{align}
In the process, we identify the primed frame indices of the sGal theory with the flavour indices of the DBI theory, that is, $\a',\b',\dotsc \leftrightarrow I,J,\dotsc$. Accordingly, for example, the map between the first elements of the lists in eq.~\eqref{map-sGal-DBI} can be written as
\begin{equation}
    \K^{\a I}_{\m}=\d^I_{\a'}\O^{\a\a'}_{\m}.
\end{equation}
Here we once again encounter the same issue concerning the mapping of spacetime indices to internal flavour ones as in the NLSM-YMS sequence. Following the same line of reasoning, our prescription is to require
\begin{equation}
     g_{IJ}=\s \d_I^{\a'}\d_J^{\b'}\eta_{\a'\b'}=-\s\d_{IJ}.\label{eq:DBI_internal_metric_identification}
\end{equation}
At the same time, we restrict ourselves to static solutions for the sGal scalar $\p$.\footnote{A naive alternative attempt to restrict to uniform (space-independent) solutions of the sGal theory would run immediately into trouble due to the antisymmetric index structure of the sGal interactions, reducing the EoM of the sGal theory to the mechanics of a free nonrelativistic particle, $\de_0^2\p=0$. This only has trivial linear solutions, and is thus of little interest.} As a consequence, the $\s=+1$ version of the sGal theory maps to the branch of the DBI theory with a negative-definite flavour metric $g_{IJ}$, corresponding to $n=D-1$ extra timelike dimensions. On the other hand, the $\s=-1$ version of the sGal theory maps to the branch of the DBI theory with $n=D-1$ extra spacelike dimensions.

We further note that, in the duality map~\eqref{map-sGal-DBI}, we simply duplicate the spin connection of the sGal theory. One copy with unprimed indices, $\mc^{\a\b}_{L\m}$, continues to play the role of the composite spin connection, whereas the other copy with primed indices accounts for the composite gauge sector of the DBI theory via 
\begin{equation}
    -\s A^{IJ}_\m= \d_{\a'}^{I}\d_{\b'}^{J} \mc^{\a'\b'}_{L\m}.
\end{equation}
The solutions of the DBI theory, obtained by a duality map from solutions of the sGal theory, therefore feature locking between the spacetime and internal gauge connections.


\subsection{Nonlinear sigma model coupled to gravity}
\label{sec:NLSMg}

The NLSM\textsubscript{g} constitutes the final entry of the duality sequence. A generic NLSM\textsubscript{g} is constructed by taking the NLSM as detailed in section~\ref{sec:NLSM} and coupling it minimally to a dynamical gravitational sector. Details of the construction can be found in section 3.4 of ref.~\cite{Brauner:2025rzv}. For the present purposes, we will however only need a version of the NLSM\textsubscript{g} with a specific choice of the coset space. To understand why, recall that in the map between the sGal and DBI theories, we duplicate the spatial geometry of the former, and turn one copy of it into an internal flavour symmetry of the latter. The same applies to the step from the DBI theory to the NLSM\textsubscript{g}. We thus introduce another set of flavour indices, for which we will use bars, that is, $\bar I,\bar J,\dotsc$. At the end of the day, we land on a NLSM\textsubscript{g} with an unbroken $\gr{SO}(n)\times\gr{SO}(m)$ flavour symmetry. The broken components of the MC form, furnishing the covariant building blocks of the matter sector of the NLSM\textsubscript{g}, accordingly carry a pair of flavour indices, one unbarred and one barred. This setup amounts to choosing the coset space of spontaneously broken internal symmetry as the real Grassmannian, $\gr{SO}(n+m)/[\gr{SO}(n)\times\gr{SO}(m)]$ or $\gr{SO}(n,m)/[\gr{SO}(n)\times\gr{SO}(m)]$~\cite{Neeling2022}. The two versions of the coset space are distinguished by the relative sign of the Euclidean metrics, $g_{IJ}$ and $g_{\bar I\bar J}$, in the two flavour spaces. The same signature corresponds to the compact coset space $\gr{SO}(n+m)/[\gr{SO}(n)\times\gr{SO}(m)]$, whereas the opposite signature corresponds to the noncompact $\gr{SO}(n,m)/[\gr{SO}(n)\times\gr{SO}(m)]$.

We can now follow verbatim the setup of section~\ref{sec:NLSM} thanks to the fact that both versions of the Grassmannian are symmetric coset spaces. The generators of the full internal symmetry group can be labelled respectively as $Q_{IJ}$ and $Q_{\bar I\bar J}$ (for the unbroken subgroup) and $Q_{I\bar I}$ (for the broken sector). Upon adding the coupling to gravity and specializing to the Grassmannian coset space, the geometric identities~\eqref{NLSM:geomids} can be  rearranged into the form
\begin{align}
\label{eq:NLSMg_Codazzi}
&\boxedcolor{Red}{\DD_\m\mc^{I\bar I}_\n=\DD_\n\mc^{I\bar I}_\m}&&\text{(Codazzi)},\\
\label{eq:NLSMg_Ricci}
&\boxedcolor{Red}{F^{IJ}_{\phantom{IJ}\m\n}=-g_{\bar I\bar J}(\mc^{I \bar I}_\m\mc^{J\bar J}_\n-\mc^{I\bar I}_\n\mc^{J\bar J}_\m)}&&\text{(Ricci)},\\
\label{eq:NLSMg_Riccibar}
&\boxedcolor{Red}{F^{\bar I\bar J}_{\phantom{\bar I\bar J}\m\n}=-g_{IJ}(\mc^{I\bar I}_\m\mc^{J\bar J}_\n-\mc^{I\bar I}_\n\mc^{J\bar J}_\m)}&&(\overline{\text{Ricci}}).
\end{align}
Here, the doubly covariant derivative is defined as
\begin{equation}
\DD_\m\mc^{I\bar I}_\n=\nabla_\m\mc^{I\bar I}_\n-g_{JK}A^{IJ}_\m\mc^{K\bar I}_\n-g_{\bar J\bar K}A^{\bar I\bar J}_\m\mc^{I\bar K}_\n,
\label{Grassmann:covder}
\end{equation}
adding a minimal coupling to the gravity sector to the gauge-covariant derivative in eq.~\eqref{NLSM:GDmcdef}.

The dynamics of the NLSM\textsubscript{g} on the Grassmannian is governed by an action, obtained from that of the NLSM by adding a minimally-coupled gravitational sector represented by the dynamical spacetime metric $g_{\m\n}$,
\begin{equation}
S_\text{NLSM\textsubscript{g}}=\int\dd^D\!x\,\sqrt{\abs{\det g}}\left(-\frac{R}{16\pi G_\text{N}}+\tfrac12\k g_{IJ}g_{\bar I \bar J}g^{\m\n}\mc^{I\bar I}_\m\mc^{J \bar J}_\n\right),
\label{NLSMg:action}
\end{equation}
where $R$ denotes as usual the Ricci scalar and $\k$ is a thus far undetermined parameter that controls the strength of the coupling between the matter and gravity sectors.\footnote{The minus sign in front of the first term, needed for positivity of the energy density of metric perturbations within the mostly-minus convention for the flat Minkowski metric, was missing in ref.~\cite{Brauner:2025rzv}; see eq.~(3.20) therein. The same remark applies to the ensuing Einstein equations~(3.21) and~(6.7). The sign discrepancy disappears once we fix the value of $\k$ by requiring duality with the DBI theory.} The NLSM\textsubscript{g} has two primary EoMs, one for the matter consisting of the NG bosons of spontaneously broken internal flavour symmetry, and one for the gravity itself,
\begin{equation}
\begin{aligned}
&\boxedcolor{Red}{\DD^\m\mc^{I\bar I}_{\m}=0}&&\text{(NLSM\textsubscript{g} matter)},\\
&\boxedcolor{Red}{R_{\m\n}=8\pi G_\text{N}\k g_{IJ}g_{\bar I\bar J}\mc^{I\bar I}_\m\mc^{J\bar J}_\n}\qquad&&\text{(Einstein)}.
\end{aligned}
\label{NLSMg:EoMscalar}
\end{equation}

Once reformulated in terms of their respective geometric data, the duality between the DBI theory and the NLSM\textsubscript{g} becomes manifest. The precise identification of the matter, gauge and gravitational sectors is as follows,
\begin{align}
    \text{DBI:}\,\,
    \{\K^{\a I}_{\m},A^{IJ}_{\m},\mc^{\a\b}_{L\m},\E^\a_\m\}
    \longrightarrow
    \text{NLSM\textsubscript{g}}:\,\,
    \{\mc^{I\bar I}_{\m},A^{IJ}_{\m},A^{\bar I\bar J}_{\m},e^\a_\m\},
    \label{map-DBI-NLSMg}
\end{align}
where $e^\a_\m$ is the dynamical coframe of the NLSM\textsubscript{g} that determines the metric via $g_{\m\n}=e^\a_\m\eta_{\a\b} e^\b_\n$. In the process, we restrict as before to static solutions of the DBI theory, and replace as appropriate frame indices with barred flavour indices of the NLSM\textsubscript{g}. Consistency between the Ricci relation~\eqref{eq:DBI_Ricci} in the DBI theory and the unbarred Ricci relation~\eqref{eq:NLSMg_Ricci} in the NLSM\textsubscript{g} then requires
\begin{equation}
g_{\bar I\bar J}
=-\d_{\bar I}^{\a}\d_{\bar J}^{\b}\eta_{\a\b},\label{eq:DBI-NLSMg-gbar}
\end{equation}
where only spatial frame indices are involved. This implies that $g_{\bar I\bar J}$ is positive-definite. Moreover, matching the Einstein equations, where we identify the flavour metrics $g_{IJ}$ on both sides of the duality, requires fixing the parameter $\k$ to
\begin{equation}
\k=-\frac{1}{8\pi G_\text{N}}. \label{eq:fix-kappa}
\end{equation}
Positivity of the scalar kinetic term in the NLSM\textsubscript{g} action~\eqref{NLSMg:action} then requires the unbarred and barred flavour metrics to have the opposite signature. Since $g_{\bar I\bar J}$ was already found to be positive-definite, the flavour metric $g_{IJ}$ should be negative-definite. We conclude that consistency of the duality map and the positivity bound on the spectrum of excitations above the NLSM\textsubscript{g} ground state require that the source theory be taken as the version of DBI with $n$ extra timelike dimensions. The corresponding target theory lives on the noncompact Grassmannian, $\gr{SO}(n,D-1)/[\gr{SO}(n)\times\gr{SO}(D-1)]$.

Another aspect of the duality map~\eqref{map-DBI-NLSMg} is that the barred gauge connection of the NLSM\textsubscript{g} descends from the spin connection in the DBI theory. The validity of the barred Ricci relation~\eqref{eq:NLSMg_Riccibar} is thus ensured by the Gauss relation~\eqref{eq:DBI_Gauss} in the DBI theory. Likewise, the Codazzi relations~\eqref{eq:DBI_Codazzi} and~\eqref{eq:NLSMg_Codazzi} map to one another without further constraints. Finally, the matter EoM in eq.~\eqref{NLSMg:EoMscalar} naturally arises from the covariant conservation law for the DBI extrinsic curvature in eq.~\eqref{DBI:EoMscalar}. Altogether, we have thus succeeded in reproducing all the dynamical EoMs and geometric identities of the NLSM\textsubscript{g}. It is worth noting that similarly to the other duality maps already discussed, also here we end up with solutions of the target theory that satisfy additional constraints not intrinsic to it. In other words, the duality map only yields a specific subclass of solutions of the NLSM\textsubscript{g}. In that regard, the identification of the barred gauge connection of the NLSM\textsubscript{g} with the spin connection of the DBI theory (also shared by the NLSM\textsubscript{g} itself) leads to a curvature constraint,
\begin{equation}
R^{\a\b}_{\phantom{\a\b}\m\n}=
\d_{\bar I}^\a \d_{\bar J}^\b F^{\bar I\bar J}_{\phantom{\bar I\bar J}\m\n}.\label{eq:NLSMg-curvature-constraint}
\end{equation}
In the same spirit, the primary DBI field equation~\eqref{eq:DBI_minimal surface} under the map~\eqref{map-DBI-NLSMg} implies a constraint satisfied by the solutions of the NLSM\textsubscript{g} obtainable by the duality map,
\begin{equation}
e^\m_\a\d^\a_{\bar I}\mc^{I\bar I}_\m = \E^\m_\a \K^{\a I}_\m = 0.
\label{eq:NLSMg_trace_constraint}
\end{equation}

We now briefly summarize the big picture of the duality for exact classical solutions of the sGal and DBI theories and the NLSM\textsubscript{g}. One fairly general possibility is to start with static solutions of the $n$-flavour DBI theory, that is, the DBI theory with $n$ extra timelike dimensions. Such solutions are mapped straightforwardly to certain solutions of the NLSM\textsubscript{g} on the noncompact Grassmannian $\gr{SO}(n,D-1)/[\gr{SO}(n)\times\gr{SO}(D-1)]$. A particularly striking special case is that of the single-flavour DBI theory, whose solutions correspond to $(D-1)$-dimensional spacelike maximal surfaces in a flat ambient Minkowski space of $D$ dimensions. These map to static solutions of the NLSM\textsubscript{g} on $\gr{SO}(1,D-1)/\gr{SO}(D-1)\simeq H^{D-1}$ in $D$ spacetime dimensions.

Alternatively, we can start from solutions of the sGal theory and map them along the entire duality chain, first to solutions of the DBI theory and subsequently to solutions of the NLSM\textsubscript{g}. Choosing the version of the sGal theory with $\s=+1$ with $D-1$ spatial dimensions, its static solutions are thus mapped to solutions of the DBI theory in $D-1$ extra timelike dimensions. These in turn map to static solutions of the NLSM\textsubscript{g} on the noncompact Grassmannian $\gr{SO}(D-1,D-1)/\bigl[\gr{SO}(D-1)\times\gr{SO}(D-1)\bigr]$. Should we instead choose the $\s=-1$ version of the sGal theory, its static solutions map to solutions of the DBI theory with $D-1$ extra spacelike dimensions. These map further onto solutions of the unphysical NLSM\textsubscript{g} with a negative-definite kinetic term for the NG fields and the compact coset space $\gr{SO}(2D-2)/\bigl[\gr{SO}(D-1)\times \gr{SO}(D-1)\bigr]$.


\subsection{Direct identification of Nambu-Goldstone fields}
\label{sec:direct-identification}

So far, we have focused exclusively on mapping the geometric data of the theories related by geometric duality. As an alternative, it turns out to be possible to connect directly the fundamental scalar degrees of freedom in these theories. The map takes a remarkably simple form provided the scalar fields are chosen judiciously.

We start with the relation between the sGal and DBI theories, again referring to ref.~\cite{Brauner:2025rzv} for details. In case of the sGal theory, there is a standard choice of the scalar $\p$ in terms of which the composite metric and the composite scalar $\X$ entering the primary field equation assume the form
\begin{align}
\label{sGal:metric}
G^{\text{sGal}}_{\m\n}&=\eta_{\m\n}-\s\de_\m\de\p\cdot\de\de_\n\p,\\
\label{sGal:primary_EoM}
\X&=\frac1{\sqrt\s}\tanh\left[\frac1D\tr\arctanh(\sqrt\s\de\de\p)\right]. 
\end{align}
Similarly, in the DBI theory, one can choose a set of flavoured scalar fields $\p^I$ so that the induced metric on the brane and the Euler-Lagrange operator defining the EoM are
\begin{align}
\label{DBI:metric}
G^{\text{DBI}}_{\m\n}&=\eta_{\m\n}-g_{IJ}\de_\m\p^I\de_\n\p^J,\\
\label{DBI:primary_EoM}
\EL^I_{\text{DBI}}&=G_{\text{DBI}}^{\m\n}\de_\m\de_\n\p^I.
\end{align}
Note that the latter is ambiguous with respect to multiplication by an arbitrary invertible matrix in the flavour space; we chose a form that will make the relation to the sGal theory manifest.

We will now show that static solutions of the sGal theory can be mapped to static solutions of the DBI theory with $D-1$ flavours of scalars by setting\footnote{The same map appeared previously in ref.~\cite{Novotny2017a} in a geometric embedded brane formulation of the sGal theory. Therein, however, it played the role of a geometric constraint, and was not used to connect solutions of the sGal theory to those of the DBI theory.}
\begin{equation}
\p^I=\d^I_i\de^i\p.
\label{eq:DBI-ansatz}
\end{equation}
In view of eq.~\eqref{eq:DBI_internal_metric_identification}, this obviously turns the metric~\eqref{sGal:metric} of the sGal theory into the metric~\eqref{DBI:metric} of the DBI theory. However, what we really need to show is that for any static solution of the sGal EoM $\X=0$, the set of fields defined by eq.~\eqref{eq:DBI-ansatz} constitutes a solution of the DBI theory. To that end, we insert the ansatz~\eqref{eq:DBI-ansatz} in eq.~\eqref{DBI:primary_EoM}, which gives
\begin{equation}
\EL^I_{\text{DBI}}=\d^I_k G_{\text{DBI}}^{ij}\de_i\de_j\de^k\p=\frac{\d^I_k}{\sqrt{\s}}\de^k\tr\arctanh\left(\sqrt{\s}\de\de\p\right),\label{DBI:primary_EoM_substitution}
\end{equation}
where we used that $\dd(\arctanh x)/\dd x=1/(1-x^2)$. With the help of eq.~\eqref{sGal:primary_EoM}, we see that vanishing of $\EL^I_{\text{DBI}}$ is equivalent to the derivative of the primary EoM $\X=0$ of the sGal theory. This proves that, indeed, any static solution of the sGal theory maps via eq.~\eqref{eq:DBI-ansatz} to a static solution of the DBI theory.

Interestingly, one can also identify, off-shell, the nonlinearly realized symmetries of the sGal and DBI theories. First of all, Galileon theories are by construction invariant under coordinate-dependent shifts $\d\p=a+b_\m x^\m$, where $a$ is a constant scalar and $b_\m$ a constant vector. Upon taking a derivative and using eq.~\eqref{eq:DBI-ansatz}, this turns to $\d\p^I=b^I$, which indeed is a symmetry of the DBI theory. The sGal theory is more prominently also invariant under
\begin{equation}
\d\p=\tfrac12s^{\m\n}(-x_\m x_\n+\s\de_\m\p\de_\n\p),\label{eq:sGal_shift}
\end{equation}
where $s^{\m\n}$ is a constant traceless symmetric matrix of parameters. Upon restricting to the spatial part, $s^{ij}$, and taking a derivative, this leads using eqs.~\eqref{eq:DBI_internal_metric_identification} and~\eqref{eq:DBI-ansatz} to the following shift of the corresponding DBI field $\p^I$,
\begin{equation}
\d\p^I=c^I_ix^i-g_{JK}c^{Ji}\p^K\de_i\p^I,
\label{DBI:nonlinsymm}
\end{equation}
where $c^I_i\equiv\d^{Ij}s_{ij}$. This is again an actual symmetry of the DBI theory.

In a similar vein, there is an elegant and direct mapping between the scalar fields of the DBI theory and those of the NLSM\textsubscript{g}. To establish this map, we need to pick a specific coset representative for the Grassmannian coset space of the NLSM\textsubscript{g}. We use the following parametrisation of the coset space, defined by
\begin{equation}
U(\pi)=\begin{pmatrix}
(\un+\epsilon\pi\pi^T)^{-1/2} & \pi(\un+\epsilon\pi^T\pi)^{-1/2}\\
-\epsilon\pi^T(\un+\epsilon\pi\pi^T)^{-1/2} & (\un+\epsilon\pi^T\pi)^{-1/2}
\end{pmatrix},
\label{eq:GS_representative_holo}
\end{equation}
where the matrix $\pi$ collects the NG fields $\pi^{I\bar I}$ and $\epsilon=\text{sign}(g_{IJ})\text{sign}(g_{\bar I\bar J})$ denotes the relative sign of the two definite flavour metrics. For $\epsilon=+1$, eq.~\eqref{eq:GS_representative_holo} provides a parametrisation of the compact Grassmannian, whereas for $\epsilon=-1$ it corresponds to the noncompact one. We will show that in parametrisation~\eqref{eq:GS_representative_holo}, static solutions of the DBI theory can be mapped to solutions of the NLSM\textsubscript{g} by setting
\begin{equation}
\pi^{I\bar I}=\d^{\bar I}_i\de^i \p^{I}.
\label{NLSMg-DBI-scalar}
\end{equation}

As the first step, we recall that in the geometric duality, the dynamical metric of the NLSM\textsubscript{g} is to be identified with the composite metric of the DBI theory. With the identification~\eqref{NLSMg-DBI-scalar} of static fields, this requires using the spatial metric
\begin{equation}
g_{ij}=G^{\text{DBI}}_{ij}=\eta_{ij}-g_{IJ} \d_{\bar I i} \d_{\bar J j} \pi^{I\bar I}\pi^{J\bar J}.
\label{metric_identification:NLSMg-DBI}
\end{equation}
For static configurations, the Euler-Lagrange operator for the NLSM\textsubscript{g} pion fields takes the form
\begin{equation}
\EL_{\text{NLSM}\textsubscript{g}}^{I\bar I}=g^{ij}\left(\nabla_i \de_j\pi^{I \bar I}-2g_{\bar J\bar K}g_{MN}\nabla_{i}\pi^{I \bar J}\left[(\un+\epsilon\pi^T \pi)^{-1}\right]^{\bar K}_{\bar M}\pi^{M\bar M}\nabla_j\pi^{N \bar I}\right).
\label{NLSMg:primary_EoM}
\end{equation}
Upon imposing the scalar identification~\eqref{NLSMg-DBI-scalar} together with the metric identification~\eqref{metric_identification:NLSMg-DBI}, this operator can be rewritten as
\begin{equation}
\begin{split}
\EL_{\text{NLSM\textsubscript{g}}}^{I\bar I}&=\d^{\bar I}_k\left(G_{\text{DBI}}^{ij}\de_i\de_j\de^k\p^I+\de^k G_{\text{DBI}}^{ij}\de_i\de_j\p^I+g_{MN}G_{\text{DBI}}^{lm}\de_m\p^M \de_l\de^k\p^I \EL_{\text{DBI}}^N \right)\\
&=\d^{\bar I}_k\left(\d^I_{\phantom{I}J}\de^k+\mathscr{A}^I_{\phantom{I}J}{}^k\right)\EL_{\text{DBI}}^J,
\end{split}
\label{scalar_EoM_map:NLSMg-DBI}
\end{equation}
where
\begin{equation}
\mathscr A^{I\phantom{J}k}_{\phantom{I}J}
\equiv g_{JL}G_{\text{DBI}}^{mn}\de_m\p^L \de_n\de^k\p^I.
\end{equation}
In addition to the scalar EoM, the NLSM\textsubscript{g} also features a separate dynamical equation for the metric, namely the Einstein equation. Its right hand side evaluates under the scalar ansatz~\eqref{NLSMg-DBI-scalar} and the metric identification~\eqref{metric_identification:NLSMg-DBI} to
\begin{equation}
\begin{split}
-g_{IJ}g_{\bar I\bar J} \left[(\un+\epsilon\pi \pi^T)^{-1}\right]^{I}_K\de_{i}\pi^{J \bar I}\left[(\un+\epsilon\pi^T \pi)^{-1}\right]^{\bar J}_{\bar K}  \de_{j}\pi^{K \bar K}=\mathcal N_{IJ}G_{\rm DBI}^{kl}\de_i\de_k\p^I\de_j\de_l\p^J,
\end{split}
\label{NLSMg:Einstein_eq}
\end{equation}
with the scalar matrix
\begin{equation}
    \mathcal N_{IJ}\equiv g_{IK}
    \left[\left(\un+\epsilon\pi\pi^T\right)^{-1}\right]^{K}_J=g_{IJ}+g_{IK}g_{JL}G_{\text{DBI}}^{mn}\de_m\p^K\de_n\p^L.
\end{equation}
Similarly, its left hand side, that is the Ricci tensor descending from the metric~\eqref{metric_identification:NLSMg-DBI}, equals
\begin{equation}
\begin{split}
    R_{ij}=-\mathcal N_{IJ}\EL_{\text{DBI}}^I \de_i\de_j\p^J+
    \mathcal N_{IJ}G_{\rm DBI}^{kl}
    \de_i\de_k\p^I\de_j\de_l\p^J,
\end{split}
\end{equation}
Together with eq.~\eqref{scalar_EoM_map:NLSMg-DBI}, we can now conclude that once the scalar fields of the DBI theory are mapped to the NG fields and metric of the NLSM\textsubscript{g} via eqs.~\eqref{NLSMg-DBI-scalar} and~\eqref{metric_identification:NLSMg-DBI}, both primary EoMs of the NLSM\textsubscript{g} are satisfied whenever $\EL_{\text{DBI}}^I$ vanishes. In other words, static solutions of the DBI theory are mapped to static solutions of the NLSM\textsubscript{g}, as we wanted to show.

In addition to the field equations, it is also interesting to relate the nonlinear symmetries of NLSM\textsubscript{g} and the DBI theory. As discussed above, the DBI theory is invariant under constant shifts of the scalar fields $\p^I$, which are seen upon using eq.~\eqref{NLSMg-DBI-scalar} to leave the pion fields $\pi^{I\bar I}$ invariant. In addition, the DBI scalars are subject to the nonlinear symmetry~\eqref{DBI:nonlinsymm}. This translates to the following transformation of the NLSM\textsubscript{g} pions,\footnote{A similar relation was already considered with a slightly different interpretation in ref.~\cite{Neeling2022}, without the possibility of mapping classical solutions and, moreover, without the inclusion of gravity.}
\begin{equation}
\d\pi^{I\bar I}=-s^{I\bar I}-g_{JK}g_{\bar J\bar K}s^{J\bar J}\pi^{K\bar I}\pi^{I\bar K}-\zeta^i\de_i\pi^{I\bar I},
\label{induced_NLSMg_transfo}
\end{equation}
where $s^{I\bar I}\equiv c^I_i\d^{i\bar I}$ and $\zeta^i\equiv g_{JK}c^{Ji}\p^{K}$. The first two terms correspond to the nonlinearly realized symmetry under $\gr{SO}(n+m)$ or $\gr{SO}(n,m)$, while the final term has the interpretation as a specific diffeomorphism. Both of these are separately symmetries of the NLSM\textsubscript{g}. Their combination featured in eq.~\eqref{induced_NLSMg_transfo} is inherited from the single nonlinear transformation~\eqref{DBI:nonlinsymm} in the DBI theory. Moreover, one can show that the variation of the NLSM\textsubscript{g} metric, expressed via eqs.~\eqref{NLSMg-DBI-scalar} and~\eqref{metric_identification:NLSMg-DBI} in terms of the DBI scalars $\p^I$, under the DBI transformation~\eqref{DBI:nonlinsymm} agrees with the Lie derivative of the metric under the infinitesimal diffeomorphism $\zeta^i$. This verifies that similarly to what we saw in case of the sGal-DBI duality, the map~\eqref{NLSMg-DBI-scalar} not only takes (static) solutions of the DBI theory to solutions of the NLSM\textsubscript{g}, but also properly embeds the off-shell symmetry of the former in that of the latter.
 

\section{Classical solutions under sGal-DBI-NLSM\textsubscript{g} duality}
\label{sec:Sol_Map2}

In this section, we illustrate the duality of solutions of the theories in the sGal-DBI-NLSM\textsubscript{g} sequence on explicit examples. We discuss three special classes of exact solutions. In section~\ref{sec:map_minimal}, we start with the single-flavour DBI theory, whose static solutions in $d$ spatial dimensions correspond to minimal or maximal surfaces in an ambient Euclidean or Minkowski space of dimension $D\equiv d+1$. Using the results of the theory of minimal surfaces in $d=2$ dimensions, we construct the geometric data and map them to the corresponding solutions of the NLSM\textsubscript{g}. In section~\ref{sec:holo}, we address the whole duality sequence, likewise in two spatial dimensions, where the exact solutions of the sGal theory can be obtained from holomorphic functions in the complex plane. In section~\ref{sec:spherically_symmetric}, we consider spherically symmetric solutions of the sGal theory, which allows us to go to higher dimensions. For both classes of solutions of the sGal theory, the corresponding solutions of the DBI theory and the NLSM\textsubscript{g} are most easily obtained using the direct scalar identification developed in section~\ref{sec:direct-identification}. Mapping the geometric data is technically more involved here, and we postpone details to appendix~\ref{app:comparison}.


\subsection{Minimal/maximal surfaces}
\label{sec:map_minimal}

Static two-dimensional solutions of the DBI theory with one extra spacelike dimension (flavour) correspond to minimal surfaces in an ambient three-dimensional Euclidean space, which admit a standard construction~\cite{Ody:1994pk}. However, as follows from the discussion in section~\ref{sec:NLSMg}, this version of the DBI theory maps to a NLSM\textsubscript{g} with a negative-definite scalar kinetic term. To reach the physical version of the NLSM\textsubscript{g}, we must instead start with the DBI theory with an extra timelike dimension. The static solutions of this theory correspond to spacelike maximal surfaces in an ambient three-dimensional Minkowski space~\cite{Kobayashi:1983max}. Below, we review the construction of both minimal surfaces in an ambient Euclidean space and maximal surfaces in an ambient Minkowski space. We will use the same sign $\epsilon$ as in the coset representative~\eqref{eq:GS_representative_holo} to distinguish the two cases, whereby $\epsilon=\pm1$ is the signature of the metric in the extra dimension; the choice $\epsilon=+1$ gives Euclidean space, whereas $\epsilon=-1$ gives Minkowski space. We will use the following notation for the line element in the ambient space and for an inner product of vectors in the ambient space,
\begin{equation}
    \dd\ell_\epsilon^2=(\dd r^1)^2+(\dd r^2)^2+\epsilon(\dd r^3)^2,
    \qquad
    \langle\vek a,\vek b\rangle_\epsilon=a^1b^1+a^2b^2+\epsilon a^3b^3,
    \label{eq:surface_ambient_metric}
\end{equation}
where $r^1,r^2,r^3$ are the components of the ambient space position vector $\vek r$. Given the restriction to a single DBI flavour, we will suppress all flavour indices $I,J,\dotsc$.

Our starting point is the local graph parametrisation of a surface in the ambient space,
\begin{equation}
    \vek r(x^1,x^2)=\begin{pmatrix}x^1\\
    x^2\\
    -\p(x^1,x^2)\end{pmatrix}.
    \label{eq:DBI-brane-coordinates}
\end{equation}
The induced metric on the surface and (the spatial part of) the corresponding DBI metric are, respectively,
\begin{equation}
    h_{ij}=\langle\de_i\vek r,\de_j\vek r\rangle_\epsilon
    =\d_{ij}+\epsilon\de_i\p\de_j\p,
    \qquad G^\text{DBI}_{ij}=-h_{ij}.
    \label{eq:surface_graph_metric}
\end{equation}
In case of $\epsilon=-1$, we impose the explicit restriction that the graph is spacelike, that is, $\d^{ij}\de_i\p\de_j\p<1$. The geometry of the surface is more conveniently described in isothermal coordinates $x^\imath\equiv(u,v)$,\footnote{Note the use of undotted indices $\imath,\jmath$ to label the isothermal coordinates. These should be distinguished from dotted indices $i,j$, which label the Cartesian coordinates $x^1,x^2$ in the ambient space.} in which
\begin{equation}
    h_{\imath\jmath}=e^{2\l(u,v)}\d_{\imath\jmath},
    \qquad G^\text{DBI}_{\imath\jmath}=-e^{2\l(u,v)}\d_{\imath\jmath}.
\end{equation}
To characterize the extrinsic geometry of the surface, we will need a normal vector. To be able to define this jointly for both signatures, we introduce the cross product $\times_\epsilon$ by $\langle\vek a\times_\epsilon\vek b,\vek c\rangle_\epsilon=\det(\vek a,\vek b,\vek c)$. The unit normal vector and the extrinsic curvature are then
\begin{equation}
    \vek n=e^{-2\l}\de_u\vek r\times_\epsilon\de_v\vek r,
    \qquad \langle\vek n,\vek n\rangle_\epsilon=\epsilon,
    \qquad \K_{\imath\jmath}=-\langle\de_\imath\vek r,\de_\jmath\vek n\rangle_\epsilon.
    \label{eq:surface_normal_curvature}
\end{equation}
Thus the normal vector has positive norm in the Euclidean case and negative norm in the Minkowski case. Reversing the normal vector flips the sign of $\K_{\imath\jmath}$ while leaving the metric unchanged.

The primary EoM of the single-flavour DBI theory for static fields $\p(x^1,x^2)$ is equivalent to the vanishing of the mean curvature,
\begin{equation}
    G_\text{DBI}^{\imath\jmath}\K_{\imath\jmath}=0.
\end{equation}
This defines a minimal surface for $\epsilon=+1$ and a spacelike maximal surface for $\epsilon=-1$. In both cases, the solutions of the condition of vanishing mean curvature admit a Weierstrass-Enneper  representation~\cite{Osserman:1969A,Dierkes:1992min,2011The,Kobayashi:1983max}. Using the complex coordinate $\xi\equiv u+\imag v$, this representation can be expressed as
\begin{equation}
    \vek r(\xi)=\frac12\operatorname{Re}\int^\xi_{\xi_0}
    \begin{pmatrix}
        1-\epsilon g(\xi')^2\\
        \imag[1+\epsilon g(\xi')^2]\\
        2g(\xi')
    \end{pmatrix}
    f(\xi')\dd\xi'.
    \label{eq:Weierstrass-Enneper-representation}
\end{equation}
The holomorphic functions $f(\xi)$ and $g(\xi)$ constitute two pieces of the Weierstrass data. With the orientation fixed above, the normal vector and extrinsic curvature are then expressed in terms of the Weierstrass data as
\begin{equation}
    \vek n=\frac1{1+\epsilon|g|^2}
    \begin{pmatrix}2\operatorname{Re}g\\2\operatorname{Im}g\\|g|^2-\epsilon\end{pmatrix},
    \qquad
    \K_{\imath\jmath}=\begin{pmatrix}
    -\operatorname{Re}(fg')&\operatorname{Im}(fg')\\
    \operatorname{Im}(fg')&\operatorname{Re}(fg')
    \end{pmatrix},
    \label{eq:surface_Weierstrass_curvature}
\end{equation}
where the prime denotes differentiation with respect to $\xi$.

Before turning to examples, let us see how both types of surfaces map to the NLSM\textsubscript{g}. For a single DBI flavour and two spatial dimensions, the target coset space is
\begin{equation}
\begin{aligned}
\gr{SO}(3)/\gr{SO}(2)&\simeq S^2\quad&\text{for }\epsilon&=+1,\\
\gr{SO}(1,2)/\gr{SO}(2)&\simeq H^2&\text{for }\epsilon&=-1,
\end{aligned}
\label{eq:surface_target_cosets}
\end{equation}
where on the second line a connected sheet is understood. The value of $\k$ fixed by duality in eq.~\eqref{eq:fix-kappa} makes the kinetic term for the NLSM\textsubscript{g} scalars positive only for $\epsilon=-1$. We nevertheless retain the compact branch ($\epsilon=+1$) as an equally valid classical-solution-generating construction.

On every patch where $x^1,x^2$ provide a nonsingular parametrisation, the direct scalar identification~\eqref{NLSMg-DBI-scalar} gives
\begin{equation}
    \pi^{\bar I}(x^1,x^2)=\d^{\bar I}_i\de^i\p(x^1,x^2)
    =-\d^{\bar I i}\de_i\p(x^1,x^2).
    \label{eq:NLSMg_scalar_minimal}
\end{equation}
Using the normal vector in eq.~\eqref{eq:surface_Weierstrass_curvature}, the same fields can be expressed directly in the isothermal coordinates as
\begin{equation}
    \pi^{\bar I}(u,v)=\frac2{1-\epsilon|g|^2}
    \bigl(\operatorname{Re}g,\operatorname{Im}g\bigr)^{\bar I}.
    \label{eq:surface_scalar_Weierstrass}
\end{equation}
For $\epsilon=+1$, the denominator can vanish at a vertical tangent plane even when the immersion is regular. For $\epsilon=-1$, the fields instead satisfy $\d_{\bar I\bar J}\pi^{\bar I}\pi^{\bar J}<1$ on every regular spacelike patch, as required for the coset representative~\eqref{eq:GS_representative_holo} to be nonsingular. The dynamical metric of the NLSM\textsubscript{g} is inherited from the DBI theory,
\begin{equation}
    g_{\imath\jmath}=G^\text{DBI}_{\imath\jmath}
    =-e^{2\l(u,v)}\d_{\imath\jmath},
    \qquad g_{00}=1,\quad g_{0\imath}=0.
    \label{eq:NLSMg_metric_minimal}
\end{equation}
This completes the data needed to uniquely specify a solution of the NLSM\textsubscript{g}.

Alternatively, one can map directly the geometric data of the surface using eq.~\eqref{map-DBI-NLSMg}. A natural isothermal coframe is $\E^\a_\imath=e^\l\d^\a_\imath$. In this frame, the corresponding MC components and connection of the NLSM\textsubscript{g} are
\begin{equation}
    \mc^{\bar I}_\imath=-e^{-\l}\d^{\bar I\jmath}\K_{\imath\jmath},
    \qquad
    A^{\bar I\bar J}_u=-\de_v\l\begin{pmatrix}0&1\\-1&0\end{pmatrix},
    \qquad
    A^{\bar I\bar J}_v=\de_u\l\begin{pmatrix}0&1\\-1&0\end{pmatrix}.
    \label{eq:NLSMg_MC_minimal}
\end{equation}
In particular, the Gauss relation~\eqref{eq:DBI_Gauss} gives
\begin{equation}
    R_{\imath\jmath}=-\epsilon\d_{\bar I\bar J}
    \mc^{\bar I}_\imath\mc^{\bar J}_\jmath,
    \qquad R_{00}=R_{0\imath}=0,
    \label{eq:surface_Einstein}
\end{equation}
which is precisely the Einstein equation of the NLSM\textsubscript{g} with the prescribed value of $\k$. The Codazzi relation~\eqref{eq:DBI_Codazzi} and vanishing mean curvature give the scalar EoM and the additional trace constraint~\eqref{eq:NLSMg_trace_constraint}.

\paragraph{Example: plane} Constant Weierstrass data, for example $g=0$ and $f=1$, generate a plane for either value of $\epsilon$. (In the Minkowski case the plane must be spacelike.) Such solutions are, however, trivial as far as the dynamical fields of the DBI theory and the NLSM\textsubscript{g} are concerned. A planar graph corresponds to an affine DBI scalar: its metric is flat and its extrinsic curvature vanishes. These solutions map to constant NG fields, vanishing MC form and a flat metric on the NLSM\textsubscript{g} side.

\paragraph{Example: catenoid} The choice $f=e^{-\xi}$ and $g=e^\xi$ gives the Euclidean catenoid and its spacelike maximal counterpart. To write both in a common notation, introduce
\begin{equation}
    c_\epsilon(u)\equiv\tfrac12(e^u+\epsilon e^{-u}),
    \qquad s_\epsilon(u)\equiv\tfrac12(e^u-\epsilon e^{-u}).
    \label{eq:surface_cs_functions}
\end{equation}
Thus $(c_+,s_+)=(\cosh u,\sinh u)$ and $(c_-,s_-)=(\sinh u,\cosh u)$, with $c_\epsilon'=s_\epsilon$, $s_\epsilon'=c_\epsilon$ and $c_\epsilon^2-s_\epsilon^2=\epsilon$. The position vector is
\begin{equation}
    \vek r=\begin{pmatrix}
    -\epsilon c_\epsilon(u)\cos v\\
    -\epsilon c_\epsilon(u)\sin v\\u
    \end{pmatrix},
    \qquad \p_\text{Cat}(u,v)=-u.
\end{equation}
The metric and extrinsic curvature are
\begin{equation}
    G^\text{DBI}_{\imath\jmath}=-c_\epsilon(u)^2\d_{\imath\jmath},
    \qquad \K_{\imath\jmath}=\operatorname{diag}(-1,1).
\end{equation}
The corresponding pion configuration in the NLSM\textsubscript{g} is
\begin{equation}
    \pi_\text{Cat}^{\bar I}(u,v)=-\frac{\epsilon}{s_\epsilon(u)}
    (\cos v,\sin v)^{\bar I}.
    \label{eq:NLSMg_scalar_Catenoid}
\end{equation}

For $\epsilon=+1$, the apparent divergence at $u=0$ reflects merely the degeneracy of the Cartesian graph projection: the induced metric and covariant MC components remain regular. For $\epsilon=-1$, the situation is different. The metric degenerates at $u=0$, where the maximal catenoid has a singular point. Its regular spacelike regions have $u\ne0$, and the norm of the pion field approaches one at the singularity. In particular, the Ricci scalar is $R=2\epsilon/c_\epsilon^4$, which diverges there on the maximal branch.

In the isothermal frame, the geometric data of the NLSM\textsubscript{g} are
\begin{equation}
    \mc^{\bar I}_u=\frac1{|c_\epsilon|}(1,0), \qquad
    \mc^{\bar I}_v=\frac1{|c_\epsilon|}(0,-1), \qquad
    A^{\bar I\bar J}_u=0, \qquad
    A^{\bar I\bar J}_v=\frac{s_\epsilon}{c_\epsilon}
    \begin{pmatrix}0&1\\-1&0\end{pmatrix}.
\end{equation}
The absolute value follows from the positive coframe factor $e^\l=|c_\epsilon|$ and matters when choosing a regular component of the maximal surface.

\paragraph{Example: helicoid} Choosing instead $f=\imag e^{-\xi}$ and $g=e^\xi$ gives the conjugate surface,
\begin{equation}
    \vek r=\begin{pmatrix}
    \epsilon s_\epsilon(u)\sin v\\
    -\epsilon s_\epsilon(u)\cos v\\-v
    \end{pmatrix},
    \qquad \p_\text{Hel}(u,v)=v.
\end{equation}
The induced metric is identical to that of the corresponding catenoid, while the extrinsic curvature differs,
\begin{equation}
    G^\text{DBI}_{\imath\jmath}=-c_\epsilon(u)^2\d_{\imath\jmath},
    \qquad \K_{\imath\jmath}=\begin{pmatrix}0&1\\1&0\end{pmatrix}.
\end{equation}
The direct scalar identification gives the following pion fields on the NLSM\textsubscript{g} side,
\begin{equation}
    \pi_\text{Hel}^{\bar I}(u,v)=-\frac{\epsilon}{s_\epsilon(u)}
    (\cos v,\sin v)^{\bar I}.
    \label{eq:NLSMg_scalar_helicoid}
\end{equation}
For $\epsilon=+1$, the graph projection again fails at $u=0$ without a singularity of the minimal immersion. For $\epsilon=-1$, the regular spacelike part has $u\ne0$. At $u=0$, the metric degenerates along a null helix and its Ricci scalar diverges. The angle $v$ is understood locally in the graph description, or on the covering space when describing the full helicoid. The geometric data of the NLSM\textsubscript{g} are
\begin{equation}
    \mc^{\bar I}_u=\frac1{|c_\epsilon|}(0,-1), \qquad
    \mc^{\bar I}_v=\frac1{|c_\epsilon|}(-1,0), \qquad
    A^{\bar I\bar J}_u=0, \qquad
    A^{\bar I\bar J}_v=\frac{s_\epsilon}{c_\epsilon}
    \begin{pmatrix}0&1\\-1&0\end{pmatrix}.
\end{equation}
For either sign, the catenoid and helicoid have the same intrinsic geometry. Their second fundamental forms differ by a $\pi/2$ rotation, and their broken MC components in the common isothermal frame are related by a constant $\gr{SO}(2)$ rotation. They are members of the associated family obtained by replacing $f$ with $e^{\imag\theta}f$, which leaves the induced metric unchanged. For $\epsilon=+1$, the plane, catenoid and helicoid are properly embedded, and only the catenoid is not simply connected.

\paragraph{Example: Enneper surface} As a final example, consider the data $f=1$ and $g=\xi^m$, where $m$ is a positive integer. For the lowest nontrivial case, $m=1$, we get
\begin{equation}
    \vek r=\frac12\begin{pmatrix}
    u-\epsilon(\tfrac13u^3-uv^2)\\
    -v-\epsilon(u^2v-\tfrac13v^3)\\
    u^2-v^2
    \end{pmatrix},
    \qquad \p_\text{Enn}=-\tfrac12(u^2-v^2).
    \label{eq:Enneper_immersion}
\end{equation}
Introducing $\rho^2\equiv u^2+v^2$ and $\ell_\epsilon\equiv1+\epsilon\rho^2$, the metric and extrinsic curvature read
\begin{equation}
    G^\text{DBI}_{\imath\jmath}=-\tfrac14\ell_\epsilon^2\d_{\imath\jmath},
    \qquad \K_{\imath\jmath}=\operatorname{diag}(-1,1).
\end{equation}
The inverse Jacobian for the Cartesian graph projection is
\begin{equation}
    J^{-1}=\frac2{\rho^4-1}\begin{pmatrix}
    -1-\epsilon u^2+\epsilon v^2&-2\epsilon uv\\
    2\epsilon uv&1-\epsilon u^2+\epsilon v^2
    \end{pmatrix},
    \qquad \det J=\tfrac14(\rho^4-1).
\end{equation}
For $\epsilon=+1$, the projection fails on the unit circle, where the tangent plane is vertical, although the induced metric is nondegenerate everywhere. The Euclidean Enneper immersion is moreover globally non-injective and self-intersects, so no single-valued graph describes the whole surface. For $\epsilon=-1$, the same unit circle is instead a genuine degeneracy of the induced metric, separating regular spacelike regions.

The direct scalar map to the NLSM\textsubscript{g} yields
\begin{equation}
    \pi_\text{Enn}^{\bar I}(u,v)=\frac2{1-\epsilon\rho^2}(u,v)^{\bar I}.
    \label{eq:NLSMg_scalar_Enneper}
\end{equation}
In the $\epsilon=+1$ case this reproduces the divergent pion field at points where the graph projection fails. In the $\epsilon=-1$ case the pion field stays finite, but its norm approaches one at $\rho=1$. This does not remove the geometric singularity where the Ricci scalar is $R=32\epsilon/\ell_\epsilon^4$. Finally, the geometric data of the NLSM\textsubscript{g} are
\begin{equation}
\begin{aligned}
    \mc^{\bar I}_u&=\frac2{|\ell_\epsilon|}(1,0), \qquad&
    \mc^{\bar I}_v&=\frac2{|\ell_\epsilon|}(0,-1),\\
    A^{\bar I\bar J}_u&=-\frac{2\epsilon v}{\ell_\epsilon}
    \begin{pmatrix}0&1\\-1&0\end{pmatrix}, \qquad&
    A^{\bar I\bar J}_v&=\frac{2\epsilon u}{\ell_\epsilon}
    \begin{pmatrix}0&1\\-1&0\end{pmatrix}.
\end{aligned}
\end{equation}
They are regular over the whole Euclidean immersion, while for $\epsilon=-1$ they define solutions only away from $\rho=1$.

\paragraph{Calabi-Bernstein theorem} The examples above illustrate a global restriction on maximal spacelike surfaces. The so-called Calabi-Bernstein theorem states that every connected spacelike maximal surface in three-dimensional Minkowski space whose induced Riemannian metric $h$ is complete is a spacelike plane~\cite{Calabi:1970bernstein,Cheng:1976max}. Thus, a nonplanar maximal surface cannot be both regular everywhere and complete. In the maximal catenoid, helicoid and Enneper cases, the singular sets lie at finite intrinsic distance from the regular regions. Removing the singularities therefore leaves an incomplete spatial metric. Allowing singularities leads to a broader class of generalized maximal surfaces~\cite{Umehara:2003max}, but the duality construction above yields regular field configurations only on their spacelike patches. Cheng and Yau extended this result to complete maximal spacelike hypersurfaces in Minkowski space of arbitrary dimension~\cite{Cheng:1976max}.


\subsection{Holomorphic solutions}
\label{sec:holo}

Next, we turn to a class of exact solutions that spans the whole sGal-DBI-NLSM\textsubscript{g} duality sequence. In two spatial dimensions, static solutions of the sGal theory can be parametrised locally by an arbitrary holomorphic function. Keeping both versions of the sGal theory as distinguished by the sign $\s=\pm1$, we map such static two-dimensional solutions first to the corresponding solutions of the two-flavour DBI theory, and subsequently to those of the NLSM\textsubscript{g}.

\paragraph{Special Galileon theory} In two spatial dimensions, the EoM of the sGal theory for static scalar fields $\p(x^1,x^2)$ boils down to the Laplace equation, $\de_i \de^i \p=0$. Its general local solution can therefore be written as the real part of an arbitrary holomorphic function, $\p(x^1,x^2)=\operatorname{Re}f(\xi)$, where $\xi\equiv x^1+\imag x^2$. The corresponding composite metric of the sGal theory is conformally flat and reads
\begin{equation}
    G_{ij}^{\gr{sGal}}=-\d_{ij}+\s \d^{kl} \de_i\de_k\p \de_l\de_j\p=-\g(\xi,\bar \xi)^2\d_{ij}, \quad \g(\xi,\bar \xi)\equiv\sqrt{1-\s \abs{f''(\xi)}^2}.
\label{eq:sGal_metric_holo}
\end{equation}
We restrict throughout to a spatial patch on which $\s\abs{f''(\xi)}^2<1$, so that the composite metric is nondegenerate and retains the required spatial signature.

\paragraph{DBI theory} The direct scalar identification~\eqref{eq:DBI-ansatz} maps the holomorphic solutions of the sGal theory to a two-flavour scalar configuration of the DBI theory, whose scalar components are given by 
\begin{equation}
\p^1(x^1,x^2)=-\operatorname{Re}f'(\xi),\qquad
\p^2(x^1,x^2)=\operatorname{Im}f'(\xi).
\label{eq:DBI_scalar_holo}
\end{equation}
The induced composite DBI metric reads
\begin{equation}
G^{\gr{DBI}}_{ij}=-\d_{ij}-g_{IJ}\de_i\p^I\de_j\p^J
=-\g(\xi,\bxi)^2\d_{ij},\label{eq:DBI_metric_holo}
\end{equation}
which naturally coincides with the composite sGal metric~\eqref{eq:sGal_metric_holo}.

\paragraph{NLSM\textsubscript{g}} We finally apply the direct scalar identification~\eqref{NLSMg-DBI-scalar} to the DBI scalars~\eqref{eq:DBI_scalar_holo}. This gives
\begin{equation}
\pi^{I\bar I}(x^1,x^2)=
\begin{pmatrix}
\operatorname{Re}f''(\xi) & -\operatorname{Im}f''(\xi)\\
-\operatorname{Im}f''(\xi) & -\operatorname{Re}f''(\xi)
\end{pmatrix}^{I\bar I}.
\label{eq:NLSMg_scalar_holo}
\end{equation}
The dynamical metric of the NLSM\textsubscript{g} is inherited from the preceding elements of the duality sequence, hence coinciding with eqs.~\eqref{eq:sGal_metric_holo} and~\eqref{eq:DBI_metric_holo}. Since the barred flavour metric is positive-definite while $g_{IJ}=-\s\delta_{IJ}$, the relative sign introduced in eq.~\eqref{eq:GS_representative_holo} is fixed by $\epsilon=-\s$. Accordingly, the $\s=+1$ branch is the physically relevant one on the NLSM\textsubscript{g} side: it passes through the DBI theory with two extra timelike dimensions and ends on the noncompact Grassmannian $\gr{SO}(2,2)/[\gr{SO}(2)\times\gr{SO}(2)]$, for which the NG fields have a positive-definite kinetic term. The $\s=-1$ branch instead passes through the DBI theory with two extra spacelike dimensions and ends with a NLSM\textsubscript{g} on the compact Grassmannian $\gr{SO}(4)/[\gr{SO}(2)\times\gr{SO}(2)]$ with a negative-definite kinetic term for the scalars.

Interestingly, the class of scalar configurations~\eqref{eq:NLSMg_scalar_holo} explores only a two-dimensional subspace of the full four-dimensional coset space of the NLSM\textsubscript{g}. A generic NG field configuration $\pi^{I\bar I}$ would be a real $2\times2$ matrix, while the matrix~\eqref{eq:NLSMg_scalar_holo}, obtained from the duality map, is in addition symmetric and traceless. In the physically more interesting case of $\s=+1$, the corresponding two broken generators together with the ``diagonal'' generator of the unbroken $\gr{SO}(2)\times\gr{SO}(2)$ close into an $\lie{so}(2,1)$ subalgebra of $\lie{so}(2,2)$. Accordingly, the image of the duality map lies in the embedded coset space $\gr{SL}(2,\R)/\gr{SO}(2)\simeq H^2$. For $\s=-1$, the analogous generators instead close into $\lie{so}(3)\subset\lie{so}(4)$, and the NG fields accordingly span the compact coset space $\gr{SU}(2)/\gr{SO}(2)\simeq S^2$.


\subsection{Spherically symmetric solutions}
\label{sec:spherically_symmetric}

For the last illustration, we consider static and spherically symmetric configurations in an arbitrary number $d\equiv D-1$ of spatial dimensions. In contrast to the previous two classes of solutions, this class allows us to go to higher dimensions, thus probing the nonlinear interactions of the sGal theory.

\paragraph{Special Galileon theory} We take $\p=f(r)$, where $r\equiv\sqrt{\d_{ij}x^i x^j}$ is the Euclidean radial distance from the origin. With this ansatz, the primary EoM of the sGal theory reduces to
\begin{equation}
(1+\sqrt\s f'')\left(r+\sqrt\s f'\right)^{d-1}=(1-\sqrt\s f'')\left(r-\sqrt\s f'\right)^{d-1},\label{radialGal_un}
\end{equation}
which can be reduced to a first order ordinary differential equation by direct integration,
\begin{equation}
\frac{1}{\sqrt{\s}}[r+\sqrt\s f'(r)]^d-\frac{1}{\sqrt{\s}}[r-\sqrt\s f'(r)]^d = c\text{(onst)}\in \R.
\label{radialGal}
\end{equation}
This can be viewed as a polynomial equation for $f'(r)$ whose order is set by the number of spatial dimensions.

For $d=1$ and $d=2$, eq.~\eqref{radialGal} is linear, reflecting the fact that the static field equation of the sGal theory in one and two spatial dimensions reduces to that of free scalars. The equation is solved by $f(r)\propto r$ and $f(r)\propto\log r$ for $d=1$ and $d=2$, respectively. The first nontrivial cases are $d=3$ and $d=4$, where the equation for $f'(r)$ is cubic.\footnote{The $d=3$ case was first studied in ref.~\cite{Curtright:2020cta}, where it was referred to as the Legendre self-dual model.} For $\s=+1$, it has exactly one real root for any real value of the constant $c$. For $d=3$ and $d=4$, respectively, the root can be expressed as
\begin{equation}
\begin{split}
    d=3: \quad & f'(r) = 2r\sinh\left[\frac{1}{3} \operatorname{arcsinh}\frac{c}{4r^3}\right],\\
    d=4: \quad &  f'(r) = \frac{2r}{\sqrt{3}}\sinh\left[\frac{1}{3} \operatorname{arcsinh}\frac{3\sqrt{3}\,c}{16r^4}\right],
\end{split}\label{eq:+d34}
\end{equation}
which can be further analytically integrated to (complicated expressions for) $f(r)$.

For $\s=-1$, the cubic equations instead admit three real distinct roots when
\begin{equation}
    \abs{c}< 4r^3\quad (d=3),\qquad
    \abs{c}<\frac{16}{3\sqrt{3}}r^4\quad (d=4).\label{intervals}
\end{equation}
Within these intervals, the three roots can be written as
\begin{equation}
\begin{split}
    d=3: \quad & f_k'(r) = 2r\sin\left[\frac{1}{3} \operatorname{arcsin}\frac{c}{4r^3}+\frac{2\pi k}{3}\right],\\
    d=4: \quad &  f_k'(r) = \frac{2r}{\sqrt{3}}\sin\left[\frac{1}{3} \operatorname{arcsin}\frac{3\sqrt{3}\,c}{16r^4}+\frac{2\pi k}{3}\right],
\end{split}\label{eq:-d34}
\end{equation}
with $k=0,1,2$. At either boundary of the intervals~\eqref{intervals}, two of the roots coalesce, while outside the intervals, the cubic equation has just one real root.

For generic $d$, a convenient parametric representation of the solutions of the EoM for $f'(r)$ can be obtained by introducing
\begin{equation}
    r=\rho\cosh(\sqrt{\s}\psi),\qquad
    f'(r)=\frac{\rho}{\sqrt{\s}}\sinh(\sqrt{\s}\psi).\label{hyperbolic_angle}
\end{equation}
The integrated radial equation~\eqref{radialGal} then reduces to $(2\rho^d/\sqrt{\s})\sinh(d\sqrt{\s}\psi)=c$. Solving for $\rho$ gives the parametric solution
\begin{equation}
    \begin{split}
    r(\psi)&=\left[\frac{\sqrt{\s}\,c}{2\sinh(d\sqrt{\s}\psi)} \right]^{1/d}\cosh\left(\sqrt{\s}\psi\right), \\
    f'(r(\psi))&=\frac{1}{\sqrt{\s}}\left[\frac{\sqrt{\s}\,c}{2\sinh(d\sqrt{\s}\psi)}\right]^{1/d}
    \sinh\left(\sqrt{\s}\psi\right),
    \end{split}\label{radialGal_para}
\end{equation}
which admits up to $2\lfloor(d-1)/2\rfloor+1$ real roots. For $\s=-1$, one finds $f'/r=\tan\psi$. Choosing $-\pi/2<\psi<\pi/2$ gives a one-to-one parametrisation of all finite real values of $f'/r$. For $\s=+1$, one instead has $f'/r=\tanh\psi$, and the parametrisation~\eqref{hyperbolic_angle} is restricted to the domain where $\abs{f'/r}<1$. The complementary region $\abs{f'/r}>1$ can be recovered by taking instead $r=\rho\sinh\psi$ and $f'(r)=\pm\rho\cosh\psi$.\footnote{Note how the parametric solution reproduces the explicit solutions in low dimensions, listed above. In case of $d=1,2$, eq.~\eqref{radialGal_para} gives $f'=c/2$ and $f'=c/(4r)$, respectively, which integrate to the expected solutions $f\propto r$ and $f\propto\log r$. For $d=3$, the $\s=+1$ branch gives $c/(4r^3)=(3x+x^3)/2$ where $x\equiv f'/r=\tanh\psi$. Substituting $x=2\sinh\theta$ and using the triple-angle identity for $\sinh3\theta$, we recover the first line of eq.~\eqref{eq:+d34}. Similarly, for $d=4$, the $\s=+1$ branch gives $3\sqrt{3}c/(16r^4)=3\sqrt{3}(x+x^3)/2$, which reproduces the second line of eq.~\eqref{eq:+d34} upon substituting $x=2/\sqrt{3} \sinh \theta$. The $\s=-1$ branch can be dealt with analogously using $x\equiv f'/r=\tan\psi$, which recovers eq.~\eqref{eq:-d34}.}

Once the radial profile $f(r)$ of the solution is fixed, the induced metric of the sGal theory follows from eq.~\eqref{sGal:metric}. Due to the spherical symmetry of the solution, it is convenient to decompose the metric into longitudinal and transverse projectors,
\begin{equation}
G_{ij}^\text{sGal}=-\d_{ij}+\s\d^{kl}\de_i\de_k\p\de_l\de_j\p=G_\parallel(r)\P_{ij}+G_\perp(r)\T_{ij},
\label{sGal:metric_spherical}
\end{equation}
where
\begin{equation}
\P_{ij}\equiv-\xx_i\xx_j,\quad
\T_{ij}\equiv-\d_{ij}+\xx_i\xx_j;\qquad
\xx^i\equiv\frac{x^i}r.
\label{sGal:PTprojectors}
\end{equation}
The coefficients of the decomposition are determined by the profile $f(r)$ via
\begin{equation}
G_{\parallel}(r)\equiv 1-\s (f'')^2,\qquad
G_{\perp}(r)\equiv 1-\s \frac{(f')^2}{r^2}.
\end{equation}

\paragraph{DBI theory} The direct scalar identification~\eqref{eq:DBI-ansatz} maps the spherically symmetric solution of the sGal theory to a $d$-flavour configuration of the DBI theory,
\begin{equation}
\p^I=\d^I_i\de^i\p=-f'(r)\d^I_i\xx^i.
\label{eq:DBI_phi2}
\end{equation}
Using this scalar configuration, one can compute all the covariant geometric quantities of the DBI theory. For instance, the induced DBI metric is expressed in terms of $\p^I$ through eq.~\eqref{DBI:metric}. It naturally coincides with the sGal metric~\eqref{sGal:metric_spherical}, once we adopt the identification~\eqref{eq:DBI_internal_metric_identification} for the flavour metric $g_{IJ}$. 

\paragraph{NLSM\textsubscript{g}} We finally apply the direct scalar identification~\eqref{NLSMg-DBI-scalar} to the DBI configuration~\eqref{eq:DBI_phi2}. The solution for the NG fields $\pi^{I\bar I}$ can be expressed in terms of either the sGal or the DBI scalars through
\begin{equation}
\pi^{I\bar I}=\d^{\bar I}_i\de^i\p^I=\d^{\bar I}_i\d^I_j\de^i\de^j\p=\frac{f'(r)}{r}\d^{I\bar I}+\left[f''(r)-\frac{f'(r)}{r}\right]\d^I_i\d^{\bar I}_j\xx^i\xx^j.
\label{eq:NLSMg_scalar_spherical}
\end{equation}
As in the holomorphic case, we keep both strands of the duality sequence and again arrive at $\epsilon=-\s$. For the physically more relevant $\s=+1$ branch, the NG fields take values from the noncompact coset space $\gr{SL}(d,\R)/\gr{SO}(d)\subset\gr{SO}(d,d)/[\gr{SO}(d)\times\gr{SO}(d)]$, whereas for the $\s=-1$ branch, they take instead values from the compact coset space $\gr{SU}(d)/\gr{SO}(d)\subset\gr{SO}(2d)/[\gr{SO}(d)\times\gr{SO}(d)]$. The general results of section~\ref{sec:direct-identification} then ensure that the NG field configuration~\eqref{eq:NLSMg_scalar_spherical}, together with the appropriate metric, inherited from the sGal and DBI theories, satisfies both the scalar EoM and the Einstein equation of the NLSM\textsubscript{g}.


\section{Higher-dimensional uplift}
\label{sec:YM}

We have outlined the different geometric duality relations that allow one to map solutions to various effective theories of NG bosons. For both duality sequences we have encountered, the final theory of the sequence includes additional propagating degrees of freedom. In case of the YMS theory these are carried by gluons, whereas in the NLSM\textsubscript{g} the extra degrees of freedom are supplied by gravity. We will now show that both duality sequences can be continued further by uplifting the solutions of the final theory to higher dimensions. In the NLSM-YMS sequence, this leads to solutions of higher-dimensional YM theory, while the sGal-DBI-NLSM\textsubscript{g} sequence admits an analogous uplift to pure Einstein gravity.


\subsection{Dimensional uplift of YMS solutions} \label{sec:YM_1}

We start with the NLSM-YMS duality sequence, and demonstrate that solutions of the YMS theory obtained from the duality map, which are subject to the relations~\eqref{YMS-condition}, can be uplifted to solutions of a higher-dimensional pure YM theory. To understand this, it is more transparent to start with the latter.


\subsubsection{General argument}

We consider a higher-dimensional Minkowski spacetime with coordinates $(x^\m,y^I)$, where $x^\m$ are the ordinary $D$-dimensional Minkowski coordinates and $y^I$ coordinates associated with additional spatial dimensions. In our mostly-minus convention, the metric in the extra dimensions is $\widehat g_{IJ}=-\d_{IJ}$. To ensure positivity of the kinetic term for the gauge fields, we assume the gauge group $G$ to be compact. We decompose the corresponding Lie algebra $\lie g$ into a direct sum $\lie h\oplus\lie m$, where $\lie h$ is a subalgebra of $\lie g$ and $\lie m$ its orthogonal complement with respect to a $G$-invariant metric on $\lie g$, and the two subspaces satisfy the conditions $[\lie h,\lie m]\subset\lie m$ and $[\lie m,\lie m]\subset\lie h$. We denote the components of the higher-dimensional gauge field belonging to the two subspaces of $\lie g$ respectively as $\widehat A^\a_{\hat\m}$ and $\widehat A^a_{\hat\m}$. The $G$-invariant metric on $\lie g$ is block-diagonal and we will denote its projections to $\lie h$ and $\lie m$ as $\widehat g_{\a\b}$ and $\widehat g_{ab}$.

Next, we demonstrate that the higher-dimensional YM theory can be dimensionally reduced to the YMS theory in $D$ dimensions. To that end, we assume a particular class of configurations of the higher-dimensional gauge field,
\begin{equation}
\widehat A^\a_{\hat\m}=(A^\a_{\m},0),\qquad
\widehat A^a_{\hat\m}=(0,\p^a_{I}),
\label{eq:YM_split}
\end{equation}
where the sole nonzero components, $A^\a_\m$ and $\p^a_I$, are independent of the extra-dimensional coordinates $y^I$. For this class of gauge field configurations, the higher-dimensional field-strength tensor, ${\widehat F}^A_{\hat\m\hat\n}=\de_{\hat\m}{\widehat A}^A_{\hat\n}-\de_{\hat\n}{\widehat A}^A_{\hat\m}+f^A_{BC}{\widehat A}_{\hat\m}^B{\widehat A}_{\hat\n}^C$, reduces to three nonvanishing blocks,
\begin{equation}
\begin{split}
F^\a_{\m\n}&=\de_\m A^\a_\n-\de_\n A^\a_\m+f^\a_{\b\g}A_\m^\b A_\n^\g,\\
F^{a}_{\m I}&=\de_\m\p^{a}_I+f^a_{\a b}A^\a_\m\p^{b}_I\equiv \Da_\m\p^{a}_I,\\
F^\a_{IJ}&=f^\a_{ab}\p^a_I\p^b_J.
\end{split}
\end{equation}
It is now easy to check that for gauge fields of the type~\eqref{eq:YM_split}, the Lagrangian of the higher-dimensional YM theory, $-\frac14\widehat g_{AB}\widehat F^A_{\hat\m\hat\n}\widehat F^{B\hat\m\hat\n}$, reduces to
\begin{equation}
\L_\text{YM}=-\tfrac14\widehat g_{\a\b}F^\a_{\m\n}F^{\b\m\n}-\tfrac12\widehat g^{IJ}\widehat g_{ab}\D_\m\p_I^{a}\D^\m\p_J^{b}-\tfrac14\widehat g^{IJ}\widehat g^{KL}\widehat g_{\a\b}f^\a_{ac}f^\b_{bd}\p_I^a\p_J^b\p_K^c\p_L^d.
\end{equation}
This agrees with our previous Lagrangian~\eqref{YMS:lagrangian} for the YMS theory provided we identify $g_{\a\b}=\widehat g_{\a\b}$ and choose the sign of $g_{IJ}$ and $g_{ab}$ so that $g^{IJ}g_{ab}=-\widehat g^{IJ}\widehat g_{ab}$. It is also easy to check that with the same identification, the higher-dimensional YM equation reduces on the ansatz~\eqref{eq:YM_split} to the EoMs~\eqref{YMS:eom} of the YMS theory. At this stage, the number of scalar flavours in the YMS theory, and hence the number of extra dimensions in the higher-dimensional YM theory, can still be arbitrary.

We will now show how the correspondence between the higher-dimensional YM theory and its dimensional reduction to the YMS theory, established by eq.~\eqref{eq:YM_split}, can be used to uplift solutions of the YMS theory, obtained via geometric duality from the NLSM. To that end, we recall that the number of YMS scalar flavours should match the number of spatial dimensions in which we solve the NLSM. In fact, the duality map between the NLSM and the YMS theory outright led us to identify $g_{IJ}=-\d_{IJ}$; cf.~eq.~\eqref{eq:YMS_flavour metric map}. Consistency with dimensional reduction of the YM theory therefore requires that
\begin{equation}
g_{ab}=-\widehat g_{ab}.
\end{equation}
Since the gauge group of the higher-dimensional YM theory was assumed to be compact, $\widehat g_{ab}$ is necessarily positive-definite, and accordingly $g_{ab}$ must be negative-definite. This is precisely the condition we found in section~\ref{sec:YMS}, needed for the duality map from the NLSM to land on a YMS theory with a positive-definite kinetic term for the scalars. It requires in turn that we start with a NLSM on a noncompact coset space.

We are now ready to spell out the advertised extended duality sequence. As explained in section~\ref{sec:YMS}, the starting point is a NLSM on a symmetric coset space $G/H$ with compact $H$ but noncompact $G$. Static solutions of such a NLSM in $d\equiv D-1$ spatial dimensions are mapped via eq.~\eqref{map-NLSM-YMS} to static solutions of the YMS theory with gauge group $H$ and a set of coloured and flavoured scalars $\p^a_I$, carrying the action of an $\gr{SO}(D-1)$ flavour symmetry. These are in turn uplifted to static solutions of pure YM theory in $2D-1$ dimensions, or equivalently solutions of Euclidean YM theory in $2D-2=2d$ dimensions. The gauge group of this YM theory is the compact version of the global symmetry group $G$ of the NLSM we started with.

The special relations~\eqref{YMS-condition} satisfied by the solutions of the YMS theory and inherited from the NLSM admit a simple reformulation in the higher-dimensional YM theory. Namely, these three conditions are equivalent to
\begin{equation}
F^\a_{ij}=\d_i^I\d_j^J F^\a_{IJ},\qquad
\d_j^JF^a_{iJ}=\d_i^IF^a_{jI},\qquad
\d^{iI}F^a_{iI}=0,
\label{eq:YM_constraints}
\end{equation}
where repeated indices imply summation over the respective coordinate block, that is, $i,j=1,\dotsc,d$ for the physical spatial coordinates and $I,J=1,\dotsc,d$ for the extra-dimensional coordinates. Note that the first relation in eq.~\eqref{eq:YM_constraints} is consistent with the first relation in eq.~\eqref{YMS-condition}. Namely, the structure constants $f^\a_{ab}$ therein differ, despite using the same symbol, by a sign, since they refer respectively to a compact gauge group and its noncompact cousin.


\subsubsection{Euclidean $\gr{SU}(2)$ YM solutions in four dimensions from NLSM}

As a concrete example, let us specialise to the case of two spatial dimensions on the NLSM side. The uplifted YM configuration then lives in a four-dimensional Euclidean space with coordinates $(x,y,u,v)$, where $(x,y)$ denote the physical spatial coordinates and $(u,v)$ the extra dimensions. For the solutions of the $\gr{SU}(1,1)/\gr{U}(1)$ NLSM, discussed in section~\ref{sec:Sol_Map1}, we obtain the following higher-dimensional gauge configuration:
\begin{equation}
\begin{aligned}
A_x&=-\frac{\imag}{\Delta}(\bar\c\c'-\c\bar\c')f^2 Q_3,\quad
&A_y&=\frac{1}{\Delta}(\bar\c\c'+\c\bar\c')f^2 Q_3,\\
A_u&=\frac{2}{\Delta}\operatorname{Im}(\mathcal W_+)Q_1-\frac{2}{\Delta}\operatorname{Re}(\mathcal W_+)Q_2,\quad
&A_v&=\frac{2}{\Delta}\operatorname{Re}(\mathcal W_-)Q_1+\frac{2}{\Delta}\operatorname{Im}(\mathcal W_-)Q_2,
\end{aligned}
\end{equation}
where $Q_A=\tau_A/2$ are the generators of $\gr{SU}(2)$ and we have used the abbreviations
\begin{equation}
\mathcal W_\pm\equiv\c'\bigl(f+|\c|^2 f'\bigr)\pm\c^2\bar\c'f',\qquad
\Delta\equiv1-|\c|^2f(|\c|^2)^2.
\end{equation}
The components of the corresponding field-strength tensor read
\begin{equation}
\begin{aligned}
F_{xy}&=F_{uv}=4\frac{|\c'|^2\left(f^2+2|\c|^2 f f'\right)}{\Delta^2}Q_3,&\\
F_{xu}&=2\operatorname{Im}(\mathcal W_{x,+})Q_1-2\operatorname{Re}(\mathcal W_{x,+})Q_2,\quad
&F_{yu}&=2\operatorname{Im}(\mathcal W_{y,+})Q_1-2\operatorname{Re}(\mathcal W_{y,+})Q_2,\\
F_{xv}&=2\operatorname{Re}(\mathcal W_{x,-})Q_1+2\operatorname{Im}(\mathcal W_{x,-})Q_2,\quad
&F_{yv}&=2\operatorname{Re}(\mathcal W_{y,-})Q_1+2\operatorname{Im}(\mathcal W_{y,-})Q_2,
\end{aligned}
\end{equation}
with
\begin{equation}
\mathcal W_{x,\pm}\equiv\D_x\left(\frac{\mathcal W_\pm}{\Delta}\right)=-\imag\mathcal W_{y,\mp}.
\end{equation}
The second equality is ensured by the profile equation~\eqref{meron:EoM1}. It guarantees that the components of the field-strength tensor in the higher-dimensional theory satisfy the relations $F_{xu}=-F_{yv}$ and $F_{xv}=F_{yu}$. Together with $F_{xy}=F_{uv}$, these imply that the entire class of static solutions of the  $\gr{SU}(1,1)/\gr{U}(1)$ NLSM uplifts to self-dual configurations of the $\gr{SU}(2)$ YM theory. This feature can be traced back to the bonus relations~\eqref{eq:YM_constraints}, which are intrinsic to the geometric formulation of the NLSM, and are thus guaranteed to hold for any solution of the higher-dimensional YM theory obtained from the NLSM via a duality map and subsequent uplift, not only for the specific solutions of the specific NLSM we have worked out explicitly. 


\subsection{Dimensional uplift of NLSM\textsubscript{g} solutions}

We next turn to the sGal-DBI-NLSM\textsubscript{g} duality sequence and ask whether it admits a higher-dimensional uplift analogous to the uplift of the NLSM-YMS sequence discussed above. As before, it is more convenient to start from the expected higher-dimensional theory.


\subsubsection{General argument}

We consider pure Einstein gravity in $D+d=1+2d$ spacetime dimensions, where $d\equiv D-1$ is the number of spatial dimensions of the lower-dimensional theory. The action reads
\begin{equation}
S_{\widehat{\mathrm{EH}}}
=-\frac{1}{16\pi \widehat G_\text{N}}\int\dd^{D+d}\!X\,\sqrt{\abs{\det\widehat g}}\,\widehat R.
\label{eq:uplift_EH_action}
\end{equation}
Next, we split the coordinates as $X^{\hat\m}=(x^\m,y^{\mathcal I})$, where ${\mathcal I}=1,\ldots,d$ labels the additional spatial directions. We restrict to a class of higher-dimensional metrics of the block-diagonal form
\begin{equation}
\dd\widehat s^2=g_{\m\n}(x)\dd x^\m\dd x^\n-M_{\mathcal I\mathcal{J}}(x)\dd y^{\mathcal I}\dd y^{\mathcal J},
\label{eq:gravity_uplift_ansatz}
\end{equation}
where the metric components are independent of $y^{\mathcal I}$, and the mixed components vanish. The matrix $M$ is by construction symmetric and positive-definite. In addition, we require
\begin{equation}
\det M=1.
\label{eq:detM}
\end{equation}
Therefore, $M$ can be thought of as parametrising the coset space $\gr{SL}(d,\R)/\gr{SO}(d)$.\footnote{In a general dimensional reduction of pure Einstein gravity, one does not need to impose any constraint on $\det M$, so the matrix $M$ parametrises the coset space $\gr{GL}(d,\mathbbm R)/\gr{SO}(d)$. The additional condition~\eqref{eq:detM} fixes the volume modulus in the extra dimensions, thus reducing the number of independent scalar degrees of freedom produced by the dimensional reduction from $d(d+1)/2$ to $d(d+1)/2-1$.} As we shall see below, for scalar configurations of the NLSM\textsubscript{g} obtained from the geometric duality, the constraint~\eqref{eq:detM} follows directly from the primary field equation of the sGal theory. Absorbing the constant coordinate volume of the extra dimensions into the $D$-dimensional Newton constant $G_\text{N}$, the reduced action now becomes
\begin{equation}
S_\text{red}
=\frac{1}{16\pi G_\text{N}}\int\dd^D\!x\,\sqrt{\abs{\det g}}
\left[
-R+\frac14\tr\left(M^{-1}\de_\m M\,M^{-1}\de^\m M\right)
\right].
\label{eq:gravity_reduced_action}
\end{equation}

To make contact with the structure of the NLSM\textsubscript{g}, we adopt the parametrisation
\begin{equation}
    M=VV^T,\qquad V\in\gr{SL}(d,\R),\label{eq:gravity_M_V}
\end{equation}
where $V$ defines an orthonormal coframe in the additional dimensions, $\widehat e^a=V_{\mathcal I}{}^a\dd y^{\mathcal I}$, with $a=1,\ldots,d$. The metric $M$ is invariant under local frame rotations $V(x)\to V(x)h(x)$, with $h(x)\in\gr{SO}(d)$. In analogy with the NLSM\textsubscript{g}, one can next introduce a ``MC form'' via
\begin{equation}
\widehat{\mathbf\O}_\m^{ab}
\equiv-\imag\delta^{ac}(V^{-1})_c{}^{\mathcal I}\de_\m V_{\mathcal I}{}^b
=-\imag\bigl(\widehat\mc_\m^{ab}+\widehat A_\m^{ab}\bigr).
\label{eq:gravity_MC_decomposition}
\end{equation}
Here, $(V^{-1})_a{}^{\mathcal I}$ denotes the dual frame, satisfying $(V^{-1})_a{}^{\mathcal I}V_{\mathcal I}{}^b=\delta_a{}^b$. The component $\widehat\mc_\m^{ab}$ is symmetric and traceless, and encodes the derivatives of the scalar degrees of freedom parametrising $\gr{SL}(d,\R)/\gr{SO}(d)$; the antisymmetric part $\widehat A_\m^{ab}$ is the corresponding composite $\gr{SO}(d)$-connection associated with the local frame redundancy. Using eq.~\eqref{eq:gravity_M_V}, we now obtain
\begin{equation}
\tr\left(M^{-1}\de_\m M\,M^{-1}\de^\m M\right)=4\delta_{ac}\delta_{bd}g^{\m\n}\widehat\mc_\m^{ab}\widehat\mc^{cd}_\n.
\label{eq:gravity_M-MC_relation}
\end{equation}
The dimensionally reduced action~\eqref{eq:gravity_reduced_action} therefore becomes
\begin{equation}
S_\text{red}=\frac{1}{16\pi G_\text{N}}\int\dd^D\!x\,\sqrt{\abs{\det g}}\bigl(-R+\delta_{ac}\delta_{bd}g^{\m\n}\widehat\mc_\m^{ab}\widehat\mc_\n^{cd}\bigr).
\label{eq:gravity_reduced_action_MC}
\end{equation}

We are now in the position to relate the dimensionally reduced version of pure gravity to the NLSM\textsubscript{g}. We will proceed in two steps: (i) match the action~\eqref{eq:gravity_reduced_action_MC} to the action~\eqref{NLSMg:action} of the NLSM\textsubscript{g}, and (ii) check that once the actions match, solutions of the EoMs of the NLSM\textsubscript{g} uplift to solutions of the higher-dimensional Einstein equation. We focus on the more physical version of the NLSM\textsubscript{g}, based on the noncompact Grassmannian. Therein, the flavour metrics were fixed to $g_{IJ}=-\d_{IJ}$ and $g_{\bar I\bar J}=+\d_{\bar I\bar J}$. In addition, the coupling $\kappa$ is required by consistency with geometric duality to take the value~\eqref{eq:fix-kappa}. To match the actions~\eqref{NLSMg:action} and~\eqref{eq:gravity_reduced_action_MC}, we define $\widehat\mc_\m^{I\bar I}\equiv\d_a^I\d_b^{\bar I}\widehat\mc_\m^{ab}$, which implies $\delta_{ac}\delta_{bd}\widehat\mc_\m^{ab}\widehat\mc_\n^{cd}=-g_{IJ}g_{\bar I\bar J}\widehat\mc_\m^{I\bar I}\widehat\mc_\n^{J\bar J}$. The two actions thus agree provided
\begin{equation}
g^{\text{red}}_{\m\n}=g^{\text{NLSM\textsubscript{g}}}_{\m\n},\qquad\widehat\mc_\m^{I\bar I}=\mc_\m^{I\bar I}.
\label{eq:gravity_reduction_matching}
\end{equation}
The spacetime metrics can be identified directly. The nontrivial condition is the matching of the MC forms, which requires reconstructing an internal metric $M_{\mathcal I\mathcal J}$ and choosing a coframe $V_\mathcal{I}{}^a$ so that the second equality holds. We next show that this condition can indeed be satisfied by the NLSM\textsubscript{g} configurations obtained from geometric duality.

Combining the direct scalar maps~\eqref{eq:DBI-ansatz} and~\eqref{NLSMg-DBI-scalar} gives
\begin{equation}
\pi^{I\bar I}=\d_i^{\bar I}\d_j^I\de^i\de^j\p.
\label{eq:gravity_pi_sGal}
\end{equation}
Thanks to the symmetry of this particular scalar field configuration in the indices $I,\bar I$, the noncompact version of the Grassmannian representative~\eqref{eq:GS_representative_holo} reduces to
\begin{equation}
U(\pi)=
\begin{pmatrix}
C_\pi&\pi C_\pi\\
\pi C_\pi&C_\pi
\end{pmatrix},
\qquad
C_\pi\equiv(\un-\pi^2)^{-1/2}.
\label{eq:Grassmannian_representative_gravity}
\end{equation}
The two combinations $C_\pi\pm\pi C_\pi$ are mutually inverse. They describe the action of $U(\pi)$ on the two subspaces of $\R^{2d}$ spanned respectively by vectors of the form $(\vek{z},\vek{z})$ and $(\vek{z},-\vek{z})$, with $\vek{z}\in\R^d$. We identify the matrix acting on the former subspace with the gravitational coframe, and that acting on the latter with its inverse,
\begin{equation}
V\equiv(\un+\pi)C_\pi=\exp(\arctanh\pi),
\qquad
V^{-1}=(\un-\pi)C_\pi=\exp(-\arctanh\pi).
\label{eq:gravity_GL_embedding}
\end{equation}
The matrix $V$ is real and symmetric, and moreover positive-definite provided the eigenvalues of $\pi$ lie in $(-1,1)$. Substitution into eq.~\eqref{eq:gravity_M_V} yields
\begin{equation}
M=VV^T=(\un+\pi)(\un-\pi)^{-1}.
\label{eq:gravity_M_pi}
\end{equation}
It is straightforward to verify that this gauge choice for the coframe ensures matching of the MC forms in eq.~\eqref{eq:gravity_reduction_matching}. To see this, we rewrite the representative~\eqref{eq:Grassmannian_representative_gravity} as
\begin{equation}
U(\pi)=\frac12
\begin{pmatrix}
V+V^{-1}&V-V^{-1}\\
V-V^{-1}&V+V^{-1}
\end{pmatrix}.
\end{equation}
A direct calculation then gives
\begin{equation}
\mathbf{\O}_\mu(\pi)=-\imag U^{-1}\de_\m U=-\frac{\imag}2
\begin{pmatrix}
V^{-1}\de_\m V-(\de_\m V)V^{-1}
&
V^{-1}\de_\m V+(\de_\m V)V^{-1}\\
V^{-1}\de_\m V+(\de_\m V)V^{-1}
&
V^{-1}\de_\m V-(\de_\m V)V^{-1}
\end{pmatrix}.
\label{eq:gravity_MC_explicit_matching}
\end{equation}
The diagonal and off-diagonal blocks are precisely the antisymmetric and symmetric parts of the gravitational MC form~\eqref{eq:gravity_MC_decomposition}, respectively. Moreover, the primary field equation of the $\s=+1$ sGal theory, which requires that $\X$ as given by eq.~\eqref{sGal:primary_EoM} vanishes, gives
\begin{equation}
\tr\arctanh\pi=0\quad\Rightarrow\quad\det V=\exp(\tr\arctanh\pi)=1\quad\Rightarrow\quad\det M=1,
\label{eq:gravity_sGal_detM}
\end{equation}
which shows that $V\in\gr{SL}(d,\R)$, reproducing precisely the condition~\eqref{eq:detM} imposed at the outset of the dimensional reduction.

It remains to show that the reconstructed metric satisfies the full higher-dimensional Einstein equation. A direct evaluation of the Ricci tensor of the metric~\eqref{eq:gravity_uplift_ansatz}, subject to the constraint $\det M=1$, gives
\begin{equation}
\begin{split}
    \widehat R_{\m\n}&=R_{\m\n}-\frac14 M^{\mathcal{I} \mathcal{J}}(\de_\m M_\mathcal{JK}) M^{\mathcal{KL}}(\de_\n M_{\mathcal{LI}})=R_{\m\n}-\d_{ac}\d_{bd}\widehat\mc_\m^{ab}\widehat\mc_\n^{cd},\\
\widehat R_{\m\mathcal I}&=0,\\
\widehat R_{\mathcal I\mathcal J}&=\frac12M_{\mathcal I\mathcal K}\nabla^\m(M^\mathcal{KL}\de_\m M_\mathcal{LJ})=\d_{ac}\d_{bd}V_\mathcal{I}{}^aV_\mathcal{J}{}^b\widehat\DD^\m\widehat\mc_\m^{cd}.\label{eq:gravity_internal_Einstein_matching}
\end{split}
\end{equation}
Here, $\nabla_\m$ denotes the Levi-Civita derivative associated with the lower-dimensional metric $g_{\m\n}$. In addition, we use eqs.~\eqref{eq:gravity_M_V} and~\eqref{eq:gravity_MC_decomposition} to obtain the relation
\begin{equation}
    M^{\mathcal K\mathcal L}\de_\m M_{\mathcal L\mathcal J}=2(V^{-1})_a{}^{\mathcal K}V_{\mathcal J}{}^b\widehat\mc_\m^{ab}.
\end{equation}
The second equality on the first line of eq.~\eqref{eq:gravity_internal_Einstein_matching} then follows by direct substitution. Similarly, evaluating the Levi-Civita covariant derivative and applying the MC decomposition~\eqref{eq:gravity_MC_decomposition} yields the second equality on the last line of eq.~\eqref{eq:gravity_internal_Einstein_matching}, with the $\gr{SO}(d)$-compatible covariant derivative defined by
\begin{equation}
    \widehat\DD_\m\widehat\mc_\n^{ab}\equiv\nabla_\mu \widehat\mc_\n^{ab}+\d_{cd}\widehat A_\m^{ac} \widehat\mc_\n^{db}+\d_{cd}\widehat A_\m^{bc} \widehat\mc_\n^{da},
\end{equation}
which coincides with eq.~\eqref{Grassmann:covder} upon the identification around eq.~\eqref{eq:gravity_reduction_matching} together with the corresponding identification of the composite connections
\begin{equation}
    A^{IJ}_\m=\d^I_a\d^J_b \widehat{A}^{ab}_\m,\qquad A^{\bar I \bar J}_\m=-\d^{\bar I}_a\d^{\bar{J}}_b \widehat{A}^{ab}_\m.
\end{equation}
The first and third expressions in eq.~\eqref{eq:gravity_internal_Einstein_matching} vanish by the EoMs~\eqref{NLSMg:EoMscalar} of the NLSM\textsubscript{g}. The mixed Ricci components vanish identically. Thus all the components of the higher-dimensional vacuum Einstein equation are satisfied. We conclude that static solutions of the NLSM\textsubscript{g} obtained via geometric duality from static solutions of the sGal theory admit an uplift to static vacuum Einstein solutions in $D+d=1+2d$ spacetime dimensions, or equivalently, after suppressing the trivial time direction, to Ricci-flat configurations of Euclidean gravity in $2d$ dimensions.

Before we conclude the general discussion of the uplift, we note that the special structure inherited from the sGal theory allows for a simple reformulation of the Euclidean uplift. Introducing the variables $u^i\equiv x^i-\d^{ij}\de_j\p$ and $v^i\equiv x^i+\d^{ij}\de_j\p$, eq.~\eqref{eq:gravity_M_pi} turns the matrix $M$ into a Jacobian,
\begin{equation}
M=\frac{\de v}{\de u}.
\label{eq:gravity_M_Jacobian}
\end{equation}
The fact that $M$ is symmetric then implies the local existence of a scalar potential $\mathcal M(u)$ such that
\begin{equation}
M_{ij}=\frac{\de^2\mathcal M}{\de u^i\de u^j}.\label{eq:M_Hessian}
\end{equation}
The condition~\eqref{eq:gravity_sGal_detM} descending from the primary field equation of the sGal theory now becomes
\begin{equation}
\det\left(\frac{\de^2\mathcal M}{\de u^i\de u^j}\right)=1,
\label{eq:gravity_Monge_Ampere}
\end{equation}
which is a special case of the (real) Monge-Amp\`ere equation.

In terms of the variables $u^i$ and $y^{\mathcal I}$, the line element of the $2d$-dimensional Euclidean gravity can be expressed as
\begin{equation}
\dd\widehat s^2=M_{ij}(u)\dd u^i\dd u^j+M_{\mathcal I\mathcal J}(u)\dd y^{\mathcal I}\dd y^{\mathcal J},
\label{eq:gravity_Euclidean_metric}
\end{equation}
where the spatial and extra-dimensional components of $M$ are identified using $M_{ij}=\d^\mathcal{I}_i\d^\mathcal{J}_j M_{\mathcal I \mathcal J}$. The Euclidean metric~\eqref{eq:gravity_Euclidean_metric} reflects algebraic constraints on the higher-dimen\-sional Riemann curvature tensor, analogous to the field-strength constraints~\eqref{eq:YM_constraints} in the uplift to the YM theory. To make these constraints explicit, we first note that the nonvanishing components of the corresponding Christoffel symbols can be expressed as
\begin{equation}
    \widehat{\Gamma}^k_{ij}=\widehat{\Gamma}^{\bar{k}}_{\bar{i}j}=\widehat{\Gamma}^{\bar{k}}_{i\bar{j}}=-\widehat{\Gamma}^k_{\bar{i}\bar{j}}=\frac12M^{kl}\de_lM_{ij}=\frac12M^{kl}\de_i\de_j\de_l\mathcal M,
\end{equation}
where the barred indices such as $\bar i=1,\dotsc,d$ run over the coordinates $y^\mathcal{I}$. With a suitably chosen orthonormal frame,
\begin{equation}
\widehat e^i=(M^{1/2})^i{}_j\,\dd u^j,
\qquad
\widehat e^{\bar i}=\d^{\bar i}_\mathcal{I}\,(M^{1/2})^\mathcal{I}{}_\mathcal{J}\,\dd y^\mathcal{J},
\label{eq:gravity_adapted_coframe}
\end{equation}
the Christoffel symbols are easily converted to the components of the spin connection. The latter in turn determines the curvature 2-form, related to the frame components of the Riemann curvature tensor via
\begin{equation}
\widehat{\mathcal R}_{\hat\a\hat\b}
\equiv
\frac12\widehat R_{\hat\a\hat\b\hat\g\hat\d}
\widehat e^{\hat\g}\wedge\widehat e^{\hat\d}.
\end{equation}
After having worked out all the details, one eventually finds that the curvature 2-form satisfies the relations
\begin{equation}
\widehat{\mathcal R}_{ij}
=\d_i^{\bar i}\d_j^{\bar j}\widehat{\mathcal R}_{\bar i\bar j},
\qquad
\d_j^{\bar j}\widehat{\mathcal R}_{i\bar j}
=\d_i^{\bar i}\widehat{\mathcal R}_{j\bar i},
\qquad\d^{i\bar j}\widehat{\mathcal R}_{i\bar j}=0,
\label{eq:gravity_constraints}
\end{equation}
which is the gravitational analogue of eq.~\eqref{eq:YM_constraints}. The first two relations are standard K\"ahler identities and restrict the curvature to the real form of $\lie{u}(d)\subset\lie{so}(2d)$, while the last follows from the sGal dynamics and removes its trace component. The resulting curvature is therefore $\lie{su}(d)$-valued. For $d=2$, eq.~\eqref{eq:gravity_constraints} reduces to the condition of (anti)self-duality.


\subsubsection{Solutions of four-dimensional gravity from holomorphic scalars}

To illustrate the uplift of NLSM\textsubscript{g} solutions to higher-dimensional pure (Euclidean) gravity, we recall the class of holomorphic solutions of the sGal theory discussed in section~\ref{sec:holo}. For the physically relevant $\s=+1$ branch, substituting the NG configuration~\eqref{eq:NLSMg_scalar_holo} into eq.~\eqref{eq:gravity_M_pi} gives
\begin{equation}
M=\frac{1}{1-\abs{f''}^2}
\begin{pmatrix}
\abs{1+f''}^2 & -2\operatorname{Im}f''\\
-2\operatorname{Im}f'' & \abs{1-f''}^2
\end{pmatrix},
\qquad \det M=1.
\label{eq:gravity_M_holo}
\end{equation}
Together with the spatial metric~\eqref{eq:sGal_metric_holo}, this yields the four-dimensional Euclidean metric
\begin{equation}
\dd\widehat s^2
=\g(\xi,\bxi)^2\d_{ij}\dd x^i\dd x^j
+M_{\mathcal{I}\mathcal{J}}\dd y^{\mathcal{I}}\dd y^{\mathcal{J}}.
\label{eq:gravity_metric_holo}
\end{equation}
As shown above, the corresponding Riemann curvature tensor is self-dual or anti-self-dual, depending on the choice of orientation.


\subsubsection{Solutions of even-dimensional gravity from isotropic scalars} 

In the same vein, we can deal with the spherically symmetric solutions of the sGal theory, discussed in section~\ref{sec:spherically_symmetric}, which allow us to construct explicit gravitational configurations in arbitrary even dimension. We focus again on the $\s=+1$ branch of the duality sequence. Substituting the NG field matrix~\eqref{eq:NLSMg_scalar_spherical}, eq.~\eqref{eq:gravity_M_pi} becomes
\begin{equation}
M_{\mathcal{I}\mathcal{J}}
=\frac{1+f''}{1-f''}\xx_\mathcal{I}\xx_\mathcal{J} +\frac{1+f'/r}{1-f'/r}\left(\d_{\mathcal{I}\mathcal{J}}-\xx_\mathcal{I}\xx_\mathcal{J}\right),
\label{eq:gravity_M_spherical}
\end{equation}
with $\xx_\mathcal{I}=\d_\mathcal{I}^i\xx_i$. The corresponding $2d$-dimensional Euclidean metric reads
\begin{equation}
\dd\widehat s^2
=\left[1-(f'')^2\right]\dd r^2
+\left[1-\frac{(f')^2}{r^2}\right]r^2\dd\Omega_{d-1}^2
+M_{\mathcal{I}\mathcal{J}}\dd y^\mathcal{I}\dd y^\mathcal{J},
\label{eq:gravity_metric_spherical}
\end{equation}
with the determinant of the metric in the extra dimensions
\begin{equation}
\det M
=\frac{1+f''}{1-f''}
\left(\frac{1+f'/r}{1-f'/r}\right)^{d-1}=1.
\label{eq:gravity_detM_spherical}
\end{equation}
Remarkably, the second equality is precisely guaranteed by the radial equation~\eqref{radialGal_un} of the sGal theory.

In the special case of $d=2$ dimensions, the integrated equation~\eqref{radialGal} reduces to
\begin{equation}
f'(r)=\frac{c}{4r},
\qquad f(r)=\frac{c}{4}\log r,
\end{equation}
up to an additive constant. Locally,
\begin{equation}
f(r)=\frac{c}{4}\log r
=\operatorname{Re}\left(\frac{c}{4}\log\xi\right),
\end{equation}
where $\xi\equiv x^1+\imag x^2$ as in section~\ref{sec:holo} and hence $r=|\xi|$. This solution is simply the rotationally invariant member of the holomorphic family discussed above. Its four-dimensional uplift satisfies the same self-duality relation.


\section{Discussion and outlook}

In this paper, we have extended the geometric duality of ref.~\cite{Brauner:2025rzv} from tree-level scattering amplitudes to exact classical configurations. The duality is made manifest by reformulating the EoMs of the different theories in terms of geometric building blocks, which descend directly from the intrinsic or extrinsic geometry of the target space of the theories. In this way, we have established two sequences of maps between exact classical solutions, relating the NLSM to the YMS theory and the sGal theory to the DBI theory and subsequently to the NLSM\textsubscript{g}. Importantly, these maps are directional, and the resulting solutions of the target theories satisfy additional geometric constraints inherited from the source theories.

One feature of the duality is worth explicitly pointing out, for it does not have an equivalent in its previous application to amplitudes. Namely, while the geometric formulation of the duality should be treated as fundamental, we found in section~\ref{sec:direct-identification} that with a suitable choice of parametrisation, it is equally possible to map directly the elementary scalar NG degrees of freedom. This provides a particularly convenient way of constructing exact solutions without having to reconstruct the geometric data at every step of the duality. In all the concrete examples of exact solutions, we were able to verify that the direct scalar map and the geometric duality are mutually compatible in that they land on the same solution of the target theory, merely expressed in terms of different variables; see appendix~\ref{app:comparison} for some details. We have not been able to prove the equivalence in full generality, mostly due to the highly nontrivial relationship between the geometric data and the elementary scalar fields. This however does not diminish the practical utility of the geometric duality as a tool for constructing new exact solutions of classical field theories. 

A further result of the present work is that both duality sequences can be extended by dimensional uplift. Solutions of the YMS theory obtained from the NLSM can be uplifted to solutions of higher-dimensional YM theory, while solutions of the NLSM\textsubscript{g}, obtained from the sGal theory, can similarly be uplifted to solutions of higher-dimensional Einstein gravity. In four Euclidean dimensions, both constructions lead to (anti-)self-dual configurations of the field strength of the YM theory and the Riemann curvature tensor of Einstein gravity, respectively. A further extension of the uplift construction, not explored in the present work, is to start from solutions of the multi-flavour DBI theory. Through the duality map, these give rise to a broader class of solutions of the NLSM\textsubscript{g} than those originating from the sGal theory. We expect that their higher-dimensional uplift would naturally involve the metric, dilaton, and antisymmetric two-form field of the Neveu-Schwarz-Neveu-Schwarz supergravity~\cite{Callan:1985ia,deWit:2002vz}. We leave the detailed construction and analysis for future work.

Just like the prequel~\cite{Brauner:2025rzv}, the present work raises some immediate questions that we would like to address in future work. The most fundamental question is what kind of field theories admit a representation in terms of a suitable set of composite operators that are invariant under reparametrisation of the elementary fields and have a clear geometric meaning. This alone would give us access to an alternative approach to the computation of scattering amplitudes, complementary to existing techniques. Perhaps even more importantly, it would be intriguing to map the landscape of physical theories to which the ideas of geometric duality might be applicable. Some immediate candidates include the Born-Infeld theory and the multi-flavour Maxwell theory coupled to gravity, both of which fit neatly in a duality tetrahedron that already accommodates all the theories discussed in this paper~\cite{Neeling2022}. Another interesting avenue for follow-up research is to explore the landscape of nonrelativistic effective theories, which are known to share many of the appealing features of their better-studied relativistic cousins and are amenable to the same on-shell methods as far as the scattering amplitudes are concerned~\cite{Mojahed2021a,Mojahed2022}. Last but not least, we stress that all the concrete, explicit solutions worked out in sections~\ref{sec:Sol_Map1} and~\ref{sec:Sol_Map2} hinged on our ability to solve the EoMs of the source theory to start with. The practical value of the geometric duality would be greatly boosted if it were possible to turn the logic and produce (at least some) on-demand solutions of the target theory.

\acknowledgments
T.W. is supported by China Scholarship Council.


\appendix

\section{Comparison of the geometric and direct scalar duality maps}
\label{app:comparison}

In this appendix, we provide some details concerning explicit checks that the duality maps based respectively on the geometric data and on the elementary scalar fields agree with each other. Such a check is relatively straightforward in case of the exact solutions of DBI theory represented by minimal surfaces (section~\ref{sec:map_minimal}). The reason for this is that for the single-flavour version of the DBI theory, there is a simple explicit general expression for the extrinsic curvature in terms of the elementary scalar field; see section~4.1 of ref.~\cite{Brauner:2025rzv} for details. We will therefore focus on the cases where the check is nontrivial. Namely, in the multi-flavour DBI theory, we do not have an explicit closed-form expression for the extrinsic curvature $\K^I_{\m\n}$ in terms of the scalars $\p^I$. Likewise, in the sGal theory, we do not have an explicit closed-form expression for the symmetric tensor $\O_{\m\n\l}$ in terms of $\p$, even though a brane construction of the sGal theory suggests that $\O_{\m\n\l}\propto\de_\m\de_\n\de_\l\p$ without any corrections of higher order in $\p$~\cite{Novotny2017a}. In these cases, in order to verify that the direct scalar map is in accord with the geometric duality, we have to reconstruct the geometric data from the available information and general consistency constraints. We demonstrate how to do so on concrete examples below.


\subsection{Holomorphic solutions}

\paragraph{Special Galileon theory} The solution for the sGal scalar $\p$ uniquely determines the metric~\eqref{eq:sGal_metric_holo}, which in turn fixes the Riemann tensor,
\begin{equation}
R_{ijkl}=-2\s\g^{-2}\bigl|f'''(\xi)\bigr|^2(\d_{ik}\d_{jl}-\d_{il}\d_{jk}).
\end{equation}
However, the geometric duality map works with objects carrying frame indices, and we therefore need to choose a (co)frame to proceed. Given the special form of the metric, a natural choice is
\begin{equation}
    \E^\a_i=\g(\xi,\bxi)\d^\a_i, \qquad\text{so that} \quad G^\text{sGal}_{ij}=
    -\d_{\a\b}\E^\a_i\E^\b_j.
    \label{eq:sGal_coframe_holo}
\end{equation}
The corresponding spin connection can then be encoded in a single complex-valued quantity,
\begin{equation}
    \mc^{\a\b}_{L\,1}+\imag \mc^{\a\b}_{L\,2}=-\imag\s\g^{-2}f''(\xi)\overline{f'''(\xi)}\begin{pmatrix}
        0 & 1 \\
        -1 & 0
    \end{pmatrix}^{\a\b}.
    \label{eq:sGal_spin_holo}
\end{equation}
The tensor $\O^{\a\a'}_\m$ can now be extracted indirectly from the Gauss relation~\eqref{eq:sGal_Gauss} and the Codazzi relation~\eqref{eq:sGal_Codazzi}. A detailed calculation shows that with the additional requirements that $\O^{\a\a'}_\m$ be traceless and the corresponding tensor $\O_{\m\n\l}$ be fully symmetric, the Gauss and Codazzi relations are sufficient to conclude that
\begin{equation}
\O^{\a\a'}_1+\imag \O^{\a\a'}_2=\g^{-2} \overline{f'''(\xi)}
    \begin{pmatrix}
        1 & -\imag \\
        -\imag & -1
    \end{pmatrix}^{\a\a'},
\label{eq:sGal_Omega_holo}
\end{equation}
up to a constant $\gr{U}(1)$ phase.

\paragraph{NLSM\textsubscript{g}} To illustrate the consistency of the geometric and direct scalar duality maps, we will now move over to the NLSM\textsubscript{g}; the DBI theory as the intermediate element of the duality sequence can be addressed in the same fashion. Note that the matrix field~\eqref{eq:NLSMg_scalar_holo} satisfies $\un_2+\epsilon\pi\pi^T=\un_2+\epsilon\pi^T\pi=\g(\xi,\bxi)^2\un_2$, where $\epsilon=-\s$. This considerably simplifies the evaluation of the components of the MC form.

To extract the individual components, we need to specify our convention for the generators. Using the notation $(E_{A}{}^B)_{CD}\equiv \d_{AC}\d^B{}_{D}$, we choose
\begin{equation}
\begin{aligned}
Q_{IJ}&=-\imag(g_{JK}E_{I}{}^{K}-g_{IK}E_{J}{}^K),\\
Q_{\bar I\bar J}&=-\imag(g_{\bar J\bar K}E_{\bar I}{}^{\bar K}-g_{\bar I\bar K}E_{\bar J}{}^{\bar K}),\\
Q_{I\bar J}&=-\imag(g_{\bar J\bar K}E_{I}{}^{\bar K}-g_{IK}E_{\bar J }{}^{K}).
\end{aligned}
\label{eq:Grassmannian_generators}
\end{equation}
Following the convention of eq.~\eqref{MCform}, this gives
\begin{equation}
\mathbf{\O}_\m
=-\imag U^{-1}\de_\m U
=-\imag\begin{pmatrix}
-g_{JK} A_\m^{IK}&\mc_\m^{I\bar J}
\\
-\epsilon\mc_\m^{J\bar I}&-g_{\bar J\bar K}A_\m^{\bar I\bar K}
\end{pmatrix}.
\label{eq:Grassmannian_MC_blocks}
\end{equation}
The rows and columns are ordered using to the unbarred and barred indices, respectively.

Upon some manipulation, we then find
\begin{align}
\notag
    \mc^{I\bar I}_1+\imag\mc^{I\bar I}_2=
    \g^{-2}\overline{f'''(\xi)}\,
    \begin{pmatrix}
        1 & -\imag \\
        -\imag & -1
    \end{pmatrix}^{I\bar I}
    &\overset{!}{=}\d^{\bar I}_{\a} \left(\K^{\a I}_1+\imag\K^{\a I}_2\right)
    =\d^I_{\a'}\d^{\bar I}_{\a} \left(\O^{\a\a'}_1+\imag \O^{\a\a'}_2\right),\label{eq:mc_NLSMg_holo} \\
    A^{IJ}_{1}+\imag A^{IJ}_2=\imag\g^{-2}f''(\xi)\overline{f'''(\xi)}\begin{pmatrix}
        0 & 1 \\
        -1 & 0
    \end{pmatrix}^{IJ} 
    &\overset{!}{=}-\s\d_{\a'}^I\d_{\b'}^J\left(\mc^{\a'\b'}_{L\,1}+\imag \mc^{\a'\b'}_{L\,2}\right),\\
\notag
    A^{\bar I\bar J}_{1}+\imag A^{\bar I\bar J}_{2}=-\imag\s\g^{-2}f''(\xi)\overline{f'''(\xi)}\begin{pmatrix}
        0 & 1 \\
        -1 & 0
    \end{pmatrix}^{\bar I \bar J}
    &\overset{!}{=}\d_{\a}^{\bar I}\d_{\b}^{\bar J} \left(\mc^{\a\b}_{L\,1}+\imag \mc^{\a\b}_{L\,2}\right).
\end{align}
Here, the expressions to the left of the symbol $\overset{!}{=}$ are obtained within the NLSM\textsubscript{g} from the NG field configuration~\eqref{eq:NLSMg_scalar_holo}. The expressions to the right of $\overset{!}{=}$ are obtained independently by transporting the geometric data in the sGal theory first to the DBI theory via eq.~\eqref{map-sGal-DBI} and then to the NLSM\textsubscript{g} using eq.~\eqref{map-DBI-NLSMg}. The two paths of the duality lead to the same results, as we wanted to demonstrate. The only remaining pieces are the metrics in the three theories, which agree by construction.


\subsection{Spherically symmetric solutions}

\paragraph{Special Galileon theory} Here, the solution for the sGal scalar $\p$ determines the composite metric via eq.~\eqref{sGal:metric_spherical}. To proceed, we again need to choose a coframe. To that end, we first introduce a shorthand notation for spatial coordinates carrying a frame index, $\xx^\a\equiv\d^\a_i\xx^i$, and use it to express the longitudinal and transverse projectors~\eqref{sGal:PTprojectors} as
\begin{equation}
    \P^\a_i\equiv-\xx^\a\xx_i,
    \qquad
    \T^\a_i\equiv\d^\a_i+\xx^\a\xx_i.
\end{equation}
The coframe can then be chosen naturally in a way consistent with the decomposition of the metric into longitudinal and transverse components,
\begin{equation}
    \E^\a_i=\sqrt{G_\parallel}\,\P^\a_i+\sqrt{G_\perp}\,\T^\a_i.
    \label{eq:sGal_coframe_spherical}
\end{equation}
The spin connection associated with this coframe is
\begin{equation}
    \mc_{L \,i}^{\a\b}=\left(\frac{\sqrt{G_{\parallel}}-\sqrt{G_{\perp}}}{r\sqrt{G_{\parallel}}}-\frac{G'_{\perp}}{2\sqrt{G_{\parallel}G_{\perp}}}\right)\left( \T^\a_i \xx^\b- \T^\b_i \xx^\a\right).
    \label{eq:sGal_spin_connection}
\end{equation}

What is less immediate is how to express the symmetric tensor $\O_{ijk}$ in terms of the sGal scalar. We proceed in two steps. First, we write down the most general static ansatz for $\O_{ijk}$, consistent with the spherical symmetry. This takes the form
\begin{equation}
\O_{ijk}=\O_{\parallel}(r)\P_{ijk}+\O_{\perp}(r)\T_{ijk},
\label{eq:sGal_form}
\end{equation}
where the fully symmetric rank-3 basis tensors are given by
\begin{equation}
    \P_{ijk}=-\xx_i \xx_j \xx_k,\qquad\T_{ijk}=\xx_k\T_{ij}+\xx_j\T_{ik}+\xx_i\T_{jk}.
\end{equation}
Second, we recover the profile functions $\O_\parallel$ and $\O_\perp$ from consistency with the Gauss relation~\eqref{eq:sGal_Gauss}. This gives, up to a common overall sign,
\begin{equation}
    \O_{\parallel}(r)= f'''(r),\qquad \O_{\perp}(r)=\frac{f''(r)}{r}-\frac{f'(r)}{r^2}.
    \label{sGal:Gauss_relation_f}
\end{equation}
It is now straightforward to check that for our static and spherically symmetric solutions of the sGal theory in $d\geq3$ spatial dimensions, $\O_{ijk}=\de_{i}\de_j\de_k\p$.

Finally, for the purposes of the duality map, we rewrite the sGal tensor in a mixed basis, with two spatial indices converted to the local frame basis,
\begin{equation}
    \O^{\a\a'}_{i}=\E^{\a j} \E^{\a'k} \O_{ijk}
        =\frac{\O_\parallel}{G_\parallel}\xx_i \P^{\a\a'}+\frac{\O_\perp}{\sqrt{G_\parallel G_\perp}}\left(\xx^\a \T_i^{\a'}+\xx^{\a'} \T_i^{\a}\right)+\frac{\O_\perp}{G_\perp}\xx_i \T^{\a\a'},\label{eq:sGal_tensor_2frame}
\end{equation}
where
\begin{equation}
    \P^{\a\a'} \equiv -\xx^\a \xx^{\a'},\qquad \T^{\a\a'} \equiv -\d^{\a\a'}+\xx^\a \xx^{\a'}.
\end{equation}

\paragraph{NLSM\textsubscript{g}} Upon adopting the parametrisation~\eqref{eq:GS_representative_holo} and evaluating the components of the MC form in terms of the NG field~\eqref{eq:NLSMg_scalar_spherical}, we find
\begin{align}
\notag
\mc^{I\bar I}_i=-\frac{\O_\parallel}{G_\parallel}\xx_i \xx^I \xx^{\bar I}+\frac{\O_\perp}{\sqrt{G_\parallel G_\perp}}\left(\xx^{\bar I}\T_i{}^{I}+\xx^{I}\T_i{}^{\bar I}\right)+\frac{\O_\perp}{G_\perp}\xx_i \T^{I\bar I}&\overset{!}{=}\d^{\bar I}_{\a}\K^{\a I}_{i}=\d^I_{\a'}\d^{\bar I}_{\a}\O^{\a\a'}_{i},\\
\label{eq:NLSMg_mc_spherical}    
    A^{IJ}_{i}=-\s\left(\frac{\sqrt{G_{\parallel}}-\sqrt{G_{\perp}}}{r\sqrt{G_{\parallel}}}-\frac{G'_{\perp}}{2\sqrt{G_{\parallel}G_{\perp}}}\right)\left( \T^I_i \xx^J- \T^J_i \xx^I\right)&\overset{!}{=}-\s\d^I_{\a'}\d^J_{\b'}\mc^{\a'\b'}_{L\,i},\\
\notag    
    A^{\bar I\bar J}_{i}=\left(\frac{\sqrt{G_{\parallel}}-\sqrt{G_{\perp}}}{r\sqrt{G_{\parallel}}}-\frac{G'_{\perp}}{2\sqrt{G_{\parallel}G_{\perp}}}\right)\left( \T^{\bar I}_i \xx^{\bar J}- \T^{\bar J}_i \xx^{\bar I}\right)&\overset{!}{=}\d^{\bar I}_{\a}\d^{\bar J}_{\b}\mc^{\a\b}_{L\,i},
\end{align}
where $\xx^I\equiv\d^I_i\xx^i$ and $\xx^{\bar I}\equiv\d^{\bar I}_i\xx^i$, and we used the projectors
\begin{equation}
\T_{i}^{I}\equiv\d_i^I+\xx_i\xx^I,\qquad
\T_{i}^{\bar I}\equiv\d_i^{\bar I}+\xx_i\xx^{\bar I},\qquad
\T^{I\bar I}\equiv-\d^{I\bar I}+\xx^I \xx^{\bar I}.
\end{equation}
Similarly to the holomorphic solutions discussed above, we have thus shown that the geometric and direct scalar duality maps are consistent with each other. One may first map the sGal scalar directly to the DBI theory and subsequently to the NLSM and only then compute the geometric data of the latter. Or one may compute the geometric data directly within the sGal theory and map them using the geometric duality. The two procedures commute with each other.


\bibliographystyle{JHEP}
\bibliography{references}

\end{document}